\documentclass{emulateapj}
\usepackage{amsmath,amssymb,epsfig,subfigure,threeparttable}
\usepackage{natbib}

\newcommand{\reffig}[1]{Figure~\ref{fig:#1}}

\newcommand\Mmin{M_{{\rm min}}}
\newcommand\Mh{M_{{\rm h}}}
\newcommand\Mhalo{\langle M_{{\rm h}} \rangle}
\newcommand\fsat{f_{\rm sat}}
\newcommand\Nc{N_{{\rm c}}(M_{{\rm h}})}
\newcommand\Ns{N_{{\rm s}}(M_{{\rm h}})}
\newcommand\sigmalogM{\sigma_{\log M}}
\newcommand\Mone{M_{1}}
\newcommand\Mzero{M_{0}}
\newcommand\Msun{M_{\odot}}
\defcitealias{bruzual2003}{BC03}
\defcitealias{ly2011}{L11}

\shorttitle
{LBGs at $z \sim 3$ in the Subaru Deep Field}
\shortauthors{Malkan et al.}

\begin{document}

\title{%
Lyman-Break Galaxies at $z\sim 3$ in the Subaru Deep Field: \\
Luminosity Function, Clustering and [OIII] Emission
%\altaffilmark{1}
}

\author{%
Matthew A. Malkan\altaffilmark{1},
Daniel P. Cohen \altaffilmark{1},
Miyoko Maruyama\altaffilmark{2},
Nobunari Kashikawa\altaffilmark{2},  \\
Chun Ly\altaffilmark{3,5},
Shogo Ishikawa \altaffilmark{2},
Kazuhiro Shimasaku\altaffilmark{2,4}, 
Masao Hayashi\altaffilmark{2}, and
Kentaro Motohara \altaffilmark{4} }

\email{malkan@astro.ucla.edu}

%\altaffiltext{1}{%
%Based on data collected at the Subaru Telescope,
%which is operated by the National Astronomical Observatory of Japan.}

\altaffiltext{1}{University of California, Los Angeles, CA 90095-1547, USA}

\altaffiltext{2}{%
Optical and Infrared Astronomy Division,
National Astronomical Observatory of Japan,
2-21-1, Osawa, Mitaka, Tokyo 181-8588, Japan}

\altaffiltext{3}{%
Observational Cosmology Laboratory, NASA Goddard Space Flight Center, 8800 Greenbelt Road, Greenbelt, MD 20771, USA
}

\altaffiltext{4}{%
 Institute of Astronomy, University of Tokyo,
2-21-1 Osawa, Mitaka, Tokyo 181-0015,
Japan}

%\altaffiltext{4}{%
%Giacconi Fellow, Space Telescope Science Institute, Baltimore, MD 21218, USA}
%\altaffiltext{7}{%
%Optical and Infrared Astronomy Division,
%National Astronomical Observatory of Japan,
%Mitaka, Tokyo 181-8588, Japan}
\altaffiltext{5}{
MMTO, Steward Observatory, Tucson, Az, 85721, USA}

%\linenumbers

%---------------------------------------------------------------------
\begin{abstract}

We combined deep $U$-band imaging from the KPNO-4m/MOSAIC camera
 with very deep multi-waveband imaging data
from optical to near-infrared wavelengths,
 to select Lyman Break Galaxies (LBGs) at $z\sim3$ using 
$U-V$ and $V-R_c$ colors in the Subaru Deep Field. 
With the resulting sample of 5161 LBGs, we construct the UV luminosity function down to $M_{\mathrm{UV}} = -18$ and find a 
steep faint-end slope of $\alpha=-1.78 \pm 0.05$. 
We %explore average stellar properties of the LBGs by analyzing 
analyze rest-frame UV-to-NIR spectral energy distributions (SEDs) generated from the median optical photometry 
and photometry on median-composite IR images (formed from stacks around the optical positions of LBGs). 
In the stacks of faint LBGs, we find a systematic background depression centered on the position of the galaxy.  
A series of experiments confirmed our view that this deficit results from the systematic difficulty of SExtractor (or any other object-finding routine) 
in finding faint galaxies in regions with higher-than-average surface densities of foreground galaxies.  
We corrected our stacked magnitudes for this effect and 
suggest that other stacking studies should also make such corrections. 
 Best-fit stellar population templates for the stacked LBG SEDs indicate stellar masses and star-formation rates 
 (SFRs) of $\log_{10}(M_*/{\rm M}_\odot) \simeq 10 $ and $ \simeq 50$ M$_\odot$ yr$^{-1}$ at $\langle i'_{\mathrm{AB}} \rangle= 24$, 
 down to $\log_{10}(M_*/{\rm M}_\odot) \simeq 8 $ and $ \simeq 3$ M$_\odot$ yr$^{-1}$ at $\langle i'_{\mathrm{AB}} \rangle= 27$. 
 For the faint stacked LBGs there is a 
 $\sim 1$ magnitude excess over the expected stellar 
continuum in the $K$-band. We interpret this excess flux as emission from
the redshifted gaseous [OIII]$\lambda\lambda 4959,5007$ and H$\beta$ lines. 
The observed excesses imply equivalent widths that increase with decreasing mass, 
reaching $\rm{EW_0([O III]4959,5007+H\beta)} \gtrsim 1500$  {\AA} (rest-frame)
in the faintest bin. 
%This result suggests that the {\it average} low-mass galaxy in the sample
 %radiates enormous power in this [OIII] doublet, exceeding $\sim 1$\% of the observed stellar light for the faintest objects. 
 Such strong [OIII] emission is seen only in a miniscule fraction of the most extreme local emission-line galaxies,
 but it probably universal in the faint galaxies that reionized the universe.
 This finding has strong implications for finding and studying the most common
 galaxies at high redshifts with upcoming surveys.
%Confirmation of this result, which will require near-infrared
%spectroscopy of these LBGs, would have significant implications for studies of cosmic reionization.
Finally, we analyze clustering by computing the angular correlation function and performing halo occupation distribution (HOD) analysis. We find a mean dark halo mass of $\log_{10}(\Mhalo/h^{-1} \Msun) = 11.29 \pm 0.12$ for the full sample of LBGs, and $\log_{10}(\Mhalo/h^{-1} \Msun) = 11.49\pm 0.1$ for the brightest half of the sample. The results support the notion that more massive dark matter halos host more luminous LBGs. 

\end{abstract}

%---------------------------------------------------------------------

%---------------------------------------------------------------------
%---------------------------------------------------------------------
\section{INTRODUCTION} \label{sec:intro}

Lyman Break Galaxies (LBGs) are selected by the strong and sharp blueward drop
in their rest-frame UV continuum, produced by the absorption of photons with energies
exceeding the intrinsic Lyman limit (rest-frame 912 {\AA}) at $z\lesssim 5.5$, or the 
Ly$\alpha$ forest of the intergalactic medium at $z \gtrsim 5.5$.  
The detection of a Lyman break requires that galaxies 
are actively star-forming, so that their stars produce a strong
blue/UV continuum.  
They therefore tend to have prominent 
young stellar populations with modest dust extinction.
The Lyman break can be defined by
optical imaging in only three bands \citep{guhathakurta1990, steidel1996}.
Thus optical LBG selection
has yielded the largest samples
of galaxies in the early universe
\citep{steidel1999,shapley2001,steidel2003,ouchi2004b,giavalisco2004,
dickinson2004,capak2004,sawicki2006,yoshida2006,iwata2007,bouwens2007,bouwens2008,bouwens2010b,bouwens2010a,
ly2009,mclure2009,fontana2010,hathi2010,castellano2010,basu2011,bielby2011,haberzettl2012,bian2013,tilvi2013}.

Much observational attention has been focused on the cosmic evolution
of the LBG luminosity function between redshifts of 1.5 and 
9 \citep{henry2007, henry2008, henry2009, mclure2010, wilkins2010,wilkins2011,bouwens2011a,bouwens2011b}.
High-redshift LBGs are relatively rare compared to the much larger
surface density of foreground objects.  Thus, not only must large fields be observed, 
but it is also useful to have more than one field to overcome the uncertainties imposed by cosmic variance.
Examples of such investigations include the European Southern Observatory $U$- and $B$-dropout survey \citep{hildebrandt2007}, and the Very Large Telescope VIMOS LBG survey \citep{bielby2011}. These surveys both combine at least five fields to cover nearly 2 deg$^2$. 
%One example of such an investigation is the ESO U- and B-dropout survey which combined 7 fields to
%cover 2 square degrees \citep{hildebrandt2007}. 
While large sky areas must be surveyed, there is a conflicting requirement for extremely deep photometry in order to study the fainter side of the LBG LF,
which is of great cosmological interest \citep{reddy2009,alavi2014}. 
Learning more about the old stellar population present in these galaxies requires deep multi-band infrared imaging over large areas, which is
even more observationally challenging than the optical imaging.
%An example of a very deep survey of U-dropouts is \citet{reddy2009}.

In this paper, we use the unique imaging of the Subaru Deep Field (SDF) [13h24m, +27$^{\circ}29'$(J2000)]
to study LBGs in a moderately wide field that is also very deep.
This provides an LBG sample with good statistics across the entire luminosity function. 
Combined with our deep complete multiwavelength photometry, we can then determine additional
LBG properties, such as stellar mass and star formation rate (SFR), and directly compare the $U$-dropout selection method with 
other search methods for high-redshift galaxies. Specifically, measurements of the stellar continuum of $z\sim 3$ galaxies at 
infrared wavelengths allow for a robust determination of stellar mass by constraining their spectral energy distributions (SEDs). 
Furthermore, optical nebular emission lines are shifted into the IR at these high redshifts; this gaseous emission can significantly 
boost broadband flux over the expected stellar continuum, changing the appearance of the SEDs \citep
{yabe2009,shim2011,atek2011,gonzalez2012,oesch2013,stark2013,deBarros2014,pirzkal2013,ly2016}. 
For example, this effect has been observed for a handful of individual $z\sim 7$ and $z\sim 8$ galaxies as an excess in the {\it Spitzer}/IRAC [3.6] 
and [4.5] $\mu$m bands, respectively, and is attributed to contamination from redshifted [OIII]$\lambda\lambda$4959,5007 
and H$\beta$ emission \citep{labbe2013,smit2014}. 
Here, we employ stacking techniques to investigate average UV-to-IR SEDs of the $z\sim 3$ LBGs down to very faint magnitudes.

A further motivation for our SDF study is the importance of measuring the spatial clustering
of a deep LBG sample in a large connected field, which
cannot be replaced by separate individual smaller fields 
\citep{giavalisco1998,giavalisco2001,foucaud2003,daddi2003,
ouchi2004a,ouchi2005,adelberger2005a,lee2006,kashikawa2006,
hildebrandt2007,quadri2007,ichikawa2007,yoshida2008,furusawa2011,bian2013,bielby2013}.
%Measuring clustering strength requires a large galaxy sample
%in a very wide, contiguous region of the sky.
% cannot probe clustering on the largest scales.
Clustering analysis of LBGs is an effective probe of
the relationship between dark matter halos and the
galaxies they host. 
%The clustering strength of distant galaxy populations can be used to infer
%the mass of hosting dark halos \citep[e.g.,][]{mo1996}.
In the standard theoretical framework of galaxy evolution,
the large-scale distribution of (visible) galaxies
is determined by the distribution of underlying (invisible) dark halos.
Increasing halo mass
is expected to be correlated with stronger spatial clustering
\citep[e.g.,][]{mo1996,nagamine2010,lacey2011}.
Such an effect has been observed in LBGs.
The clustering strength of LBGs increases with UV luminosity
\citep[e.g.,][]{giavalisco2001,ouchi2004a,adelberger2005a,
lee2006,hildebrandt2007}. Since dust-corrected UV luminosity is directly related to SFR,
this may imply that the star formation activity of LBGs
is controlled by the mass of dark halos that host them.

The very deep multi-wavelength photometry that has
been assembled in the large area of the SDF
now allows us to make a detailed survey of LBGs
and their properties and clustering.
To construct a large sample of LBGs at $z\sim 3$,
we obtained $U$-band imaging of the SDF with the MOSAIC camera
on the Mayall 4 m telescope.
This was combined with the very deep optical multi-waveband imaging data
obtained by the Suprime-Cam,  
along with deep $J$, $H$, and $K$ data taken with the wide-field
near-infrared camera (WFCAM) on the United Kingdom Infrared Telescope (UKIRT),
$H$-band imaging from the NOAO Extremely Wide-Field Infrared Imager (NEWFIRM),
and {\it Spitzer Space Telescope} Infrared Array Camera (IRAC)
maps taken at 3.6, 4.5, 5.8, and 8.0 $\mu$m.

The organization of this paper is as follows. 
In \S\ref{sec:data}, an account of the observations
and the data is presented.
The selection of the large sample of 
$z\sim 3$ LBGs is described in \S\ref{sec:lbg}.
We also describe simulations to assess the redshift distribution function
and the contamination of the sample by interlopers.
The luminosity function of the LBGs is presented in \S\ref{sec:lf}. 
We infer global physical properties of the LBGs
from their stacked UV-to-IR SEDs in
\S\ref{sec:LBGprops}. Here we also report a detection of
[OIII]$\lambda\lambda$4959,5007 + H$\beta$ emission-line 
contamination in the stacked $K$-band
data, which implies very high equivalent widths in faint, low-mass
LBGs. The clustering analysis is presented in \S\ref{sec:clustering}.
%A subset of the most massive LBGs is studied separately in the Appendix.
%These unusually luminous LBGs show stronger clustering, corresponding to more
%massive dark halos.
Our summary and conclusions are given in \S\ref{sec:sum}.

Throughout this paper, we use the AB magnitude system
\citep{oke1983}.
We adopt a flat universe cosmology with
$\mathrm{\Omega}_m = 0.3$, $\mathrm{\Omega}_\mathrm{\Lambda} = 0.7$,
and  $H_0 = 100\ h$ km s$^{-1}$ Mpc$^{-1}$ where $h=0.7$,
$\sigma_8 = 0.9$, and baryonic density $\mathrm{\Omega}_b = 0.04$.
However, all dark halo mass parameters presented in the clustering analysis are 
expressed in units of $h^{-1}$ Mpc to facilitate comparison with previous results.

%---------------------------------------------------------------------
%---------------------------------------------------------------------
\section{DATA} \label{sec:data}

\subsection{Imaging Data} \label{subsec:data-imaging}

The SDF has a set of very wide and deep
multi-waveband optical imaging data
from the Subaru Deep Survey project.\footnote[6]{The SDF images and catalogs 
are available at \href{http://soaps.nao.ac.jp/}{http://soaps.nao.ac.jp/}.}
These data cover one Subaru/Suprime-Cam field \citep{miyazaki2002}, $34'\times27'$ in size, with $0\farcs202$ pixels,
in five standard broad-band filters,
$B$, $V$, $R_c$, $i'$, and $z'$. The observations, data processing, and the source detections are described in \citet{kashikawa2004}. Basic parameters are summarized in Table \ref{table:photometry}.
The $3\sigma$ limiting magnitudes of the Suprime-Cam imaging, measured
in $2\arcsec$-diameter apertures, are $28.45$, $27.74$, $27.80$, $27.43$, $26.62$ in $B$, $V$, $R$, $i'$, and $z'$, respectively.
The final co-added image has an effective area of $876$ arcmin$^2$, with a seeing FWHM of $0.98''$ in each band.
The wide FOV enables us to probe large comoving volumes exceeding $10^6$ Mpc$^3$ for our LBG sample at $\langle z \rangle\simeq3$.
The absolute error of astrometry was found to be $0\farcs 21$-$0\farcs 27$.

Below, we describe the $U$-band, NIR ($J$-, $H$-, and $K$-band), and mid-IR data of
the SDF utilized for this study. The photometric sensitivity, seeing,
and image scales and sizes for these data are summarized
in Table \ref{table:photometry}. 

\begin{deluxetable*}{ccccccc}
%\tabletypesize{\textsize}
\tablewidth{\textwidth} 
\tablecaption{Photometry in Subaru Deep Field.
\label{table:photometry}}
\tablehead{\colhead{Band}  & \colhead{$\lambda_{\rm{cent}}$\,\tablenotemark{a}} & 
 \colhead{$\Delta \lambda$\,\tablenotemark{b}} & \colhead{3$\sigma$ Limit\,\tablenotemark{c}} &
  \colhead{Seeing FWHM} & \colhead{SDF Coverage\tablenotemark{d}} \\
\colhead{   } & \colhead{({\AA})} & \colhead{({\AA})}  & \colhead{[AB-mag, $2''$]} &
\colhead{(arcsec)}  & \colhead{\%} }

\startdata
 $U$ & 3630 & 750 &  27.19  &1.49 & 100\\ 
$B$ & 4440  & 690 & 28.45  & 0.93  & 100\\ 
$V$ &  5460 & 890 & 27.74  & 0.940  & 100\\ 
$R_c$ & 6520 & 1100 &  27.80  & 0.97  & 100 \\ 
$i'$ & 7660 & 1420 & 27.43  & 0.98  & 100\\ 
$z'$ & 9020 & 960 & 26.62  & 0.98 & 100\\ 
$J$ & 12560 & 1600 & 24.75  & 1.10 & 82 \\
$H$(WFCAM) & 16510 & 2900 & 24.05 & 1.10 & 67 \\ 
$H$(NEWFIRM) & 16190  & 3000 & 24.20 & 0.90  & 95\\ 
$K$  & 22360 & 3400 &  24.15  & 1.10 & 82 \\  
 $[3.6]$  & 35500 & 7500 & 24.29 & 1.92 &  81   \\
$[4.5]$ & 44930 & 10150   & 23.97 & 1.79 &  75  \\
 $[5.8]$ & 57310 & 14250  & 22.86  & 2.00 &  81  \\
$[8.0]$ & 78720 & 29050 & 22.89  & 2.23 &  75 \\   
\enddata

\tablenotetext{a}{Centroid wavelength for the filters in {\AA}.} 
\tablenotetext{b}{Filter bandwidths in {\AA}, not counting atmospheric transmission.}
\tablenotetext{c}{Limiting magnitudes measured in $2''$ diameter apertures across all portions of the image utilized in the study 
(across regions of non-uniform depth). } 
\tablenotetext{d}{Coverage of the SDF in each image that was utilized for this study. Full coverage corresponds to 876 arcmin$^2$. }

\end{deluxetable*}

\subsubsection{$U$-band Imaging} \label{subsec:U-band}

$U$-band data were obtained from the Kitt Peak National Observatory
Mayall 4 meter telescope using
the MOSAIC-1 Imager \citep{muller1998}, with a FoV of 36\arcmin, on 2007 April 18/19 in the SDF
under NOAO proposal Proposal ID 2007A-0589 \citep{ly2007a}. Details
of the observations and data reduction are described in \citet{ly2011} (hereafter, L11). 

The imaging was centered on the single FoV of the Suprime-Cam.
% with a pixel scale of $0\farcs26$ pixel$^{-1}$.
Observing conditions were dark (new moon) with minimal
cloud coverage. A series of 25-minute exposures was obtained to accumulate
47.4 ks of integration. A standard dithering pattern was followed to provide 
uniform imaging across the CCD gaps of about 10\arcsec.

The MSCRED IRAF (version 2.12.2) package was used to reduce the data and
produced final mosaic images with a pixel scale of 0\farcs258 and an
average-weighted seeing of $1\farcs49$ FWHM. The reduction steps closely followed
the procedures outlined for the reduction of the NOAO Deep
Wide-Field Survey MOSAIC data.

While several standard star fields \citep{landolt1983} were observed,
Landolt's $U$-band filter differs significantly from the MOSAIC/$U$, which
makes it difficult to photometrically calibrate the data. Instead, 102
Sloan Digital Sky Survey (SDSS) stars distributed uniformly across the eight
Mosaic-1 CCDs with $u\arcmin \lesssim 21$ mag were used. This approach
requires a transformation between SDSS $u$\arcmin\ and MOSAIC/$U$, which
was obtained by convolving the spectrum of 175 Gunn-Stryker stars \citep{gunn1983} with the
total system throughput at these wavebands. 
%These stars span $B-V = -0.2$
%to 0.8, and show that the $B-V$ color term is smaller than the scatter
%(1$\sigma$ = 0.06 mag) in the data and that the $U-u$\arcmin\ color has a scatter of
%0.05 mag. 
%The $3\sigma$ limiting magnitude within a 3$''$ aperture is 26.4 (AB),
%after applying appropriate aperture corrections to total magnitudes.
The 3$\sigma$ limiting $U$-magnitude is 27.19 in a $2''$-diameter aperture,
and 26.52 in a $3''$-diameter aperture.

\subsubsection{Near-infrared Imaging} \label{subsec:NIRdata}

$J$, $H$, and $K$-band imaging of the SDF was taken with UKIRT/WFCAM
\citep{casali2007} during the period from 2005 to 2010. The initial data set in $J$
and $K$, which cover a part of the SDF, and full data in $K$ are described
in \citet{hayashi2007} and \citetalias{ly2011}. Because WFCAM is
composed of four detectors with $13.65'\times13.65'$ FoVs with a gap of
$12.83'$, four pointings are required to cover the SDF continuously.
With the final data set, the
$J$ and $K$-band data cover the whole area of the SDF, with varying image depths
across the field (see below). The $H$
data are available for only 67\% of the SDF, as we 
could not complete the fourth pointing in this band.
%This is because WFCAM is
%composed of four detectors with 13.65'$\times$13.65' FoV with a gap of
%12.83' and requires four pointings to cover the SDF continuously.
%This is because WFCAM is
%composed of four detectors with $13.65'\times13.65'$ FoV with a gap of
%$12.83'$ and requires four pointings to cover the SDF continuously.
%Unfortunately, we had no opportunity of observation in the fourth
%pointing in $H$-band.
 The integration times vary depending on the
pointings in each band; 1.1$-$10 hours in $J$, 1.0$-$5.0 hours in $H$,
and 0.4$-$5.0 hours in $K$. Data reduction was made in the standard manner for NIR imaging data
using our own IRAF-based pipeline. 
%We took special care in the reduction of WFCAM data to exclude 
Spurious objects due to crosstalk
of bright objects were carefully excluded. Next, all the images were smoothed with a Gaussian
kernel so that their PSF sizes have $1.1''$ FWHM. Astrometry and flux
calibration to derive the magnitude zero point were performed by using
the 2MASS point-source catalog \citep{skrutskie2006}. The mosaic of the images taken in different pointings result in
non-uniformity in the integration times over the SDF. Nearly half
(49\%) of the SDF is uniformly deep in all three bands, with
5$\sigma$ limiting magnitudes in a $2''$ diameter aperture of 24.2 ($J$),
23.5 ($H$), and 23.6 ($K$). On the other hand, 33\% of the SDF has
moderately deep $J$ and $K$ data with limiting mags of 23.2 ($J$) and 23.4
($K$). The remaining 18\% of the SDF is covered in all bands, but only
shallow data are available with depths of 22.8 ($J$), 22.5 ($H$), and 22.0 ($K$).

We also utilize more recently acquired $H$-band data with NEWFIRM. 
The NEWFIRM imaging data were acquired on 2012 March 06/07 and 2013 March 27/30 with photometric observing conditions at KPNO, clear skies, and 0\farcs9 seeing. These conditions were significantly better than in 2008, when the seeing was 1\farcs3 with transparency varying by as much as $\approx$2 mag. The improved observing conditions yielded a mosaicked image that is $\approx$1.8 mag deeper than our previous $H$-band observations described in \citetalias{ly2011}. These NEWFIRM data were reduced following the steps outlined in \citet{ly2011a} with the \textsc{IRAF} \textit{nfextern} package.

\subsubsection{Mid-infrared Imaging} \label{subsec:MIRdata}

The SDF has been imaged in the mid-infrared with {\it Spitzer}/IRAC Channels 1$-$4, at 3.6,
4.5, 5.8, and 8.0 $\mu$m \citep{fazio2004}. The data 
used for this project were super mosaics obtained from the {\it Spitzer} Heritage
Archive\footnote{\href{http://sha.ipac.caltech.edu/applications/Spitzer/SHA/}{http://sha.ipac.caltech.edu/applications/Spitzer/SHA/}}. Contributing
observations to these mosaics are described in \citet{ota2010}, \citet{jiang2013}, 
and \citetalias{ly2011}. These mosaics cover roughly 80\% of the SDF in [3.6] and [5.8] and 75\% in [4.5] and [8.0]. 
The median PSF FWHMs are $1\farcs92$, $1\farcs79$, $2\farcs00$, and 
$2\farcs23$ in the [3.6], [4.5], [5.8], and [8.0] mosaics, respectively. The $3\sigma$ limiting magnitudes, estimated by performing photometry in $2''$ apertures placed randomly throughout the mosaics, are 24.29 ([3.6]), 23.97 ([4.5]), 22.86 ([5.8]), and 22.89 ([8.0]). Additional information on the IRAC instrument properties can be found in the IRAC Instrument Handbook.\footnote{\href{http://irsa.ipac.caltech.edu/data/SPITZER/docs/irac/iracinstrumenthandbook/}{http://irsa.ipac.caltech.edu/data/SPITZER/docs/irac/\\iracinstrumenthandbook/}}

\subsection{Photometric Catalogs} \label{subsec:Photometry}

Object detection and photometry were performed using SExtractor \citep[version 2.3;][]{bertin1996}.
We masked out regions of low S/N at the edges of the image, near bright
stellar halos, and saturated CCD blooming, as well as bad pixels of
abnormally high or low count spikes.
We first detected objects in the (extremely deep) $R_c$-band image.
We considered an object detected when if had more than five contiguous
pixels detected at greater than the $2\times \mathrm{RMS}$ threshold.

For each detected object,
photometry was performed in the other waveband images at exactly the same positions
by running SExtractor in dual-image mode.  
We adopted Kron aperture magnitudes (MAG\_AUTO) in SExtractor for total magnitudes,
and used magnitudes within a $2''$ diameter aperture to 
derive colors\footnote{We also measured magnitudes within a larger aperture of
$3''$ diameter,
as the PSF FWHM of the final images is somewhat larger in the $U$-band than in the $BVR_c i' z'$ bands.
However, because over 90 \% of the LBG candidates selected with
$2''$ aperture magnitudes overlap with those
selected with $3''$ aperture magnitudes,
we adopted $2''$ aperture magnitudes
in order to obtain the colors of faint objects with better S/N ratios.}.
We applied a differential aperture correction of $-0.2$ mag to the $U$ magnitudes
to compensate for the larger PSF size in the $U$-band image.
The value $-0.2$ was determined so that the difference between
the aperture magnitudes and the total magnitudes of the $U$ band
became equal to those of the other bands for LBG candidates.

The magnitudes of objects were corrected for a small amount of
foreground Galactic extinction using the dust map of
\citet{schlegel1998}.
The value of $E(B-V)=0.017$ corresponds to an extinction of
$A_U = 0.08$, $A_B = 0.07$, $A_V = 0.05$, $A_{R_c} = 0.04$, $A_{i'} = 0.03$,
$A_{z'} = 0.02$.
% $A_\mathrm{NB816} = 0.03$, and $A_\mathrm{NB921} = 0.02$.

%---------------------------------------------------------------------
%---------------------------------------------------------------------
\section{LYMAN-BREAK GALAXY SAMPLE AT $z\sim 3$}
\label{sec:lbg}

\subsection{Selection of Lyman-break Galaxies}
\label{subsec:lbg-select}

We find that a combination of $U$, $V$, and $R_c$ bands works best
to select LBGs at $z\sim 3$ among our bandpass set.
The $V$ and $R_c$ filters are redward of Ly$\alpha$ at $z = 2.9$, while half of the $U$-band
is entirely below the Lyman limit, causing a substantial jump
in the $U-V$ color.  We note that comparing the $V$ filter (rather than the $B$)
to find $U$-band ``drop-outs" mostly eliminates the ambiguity from galaxies
at somewhat lower redshifts, where a red color could be produced by
partial absorption due to the Ly$\alpha$ forest and intervening Lyman
limit absorbers, rather than a true Lyman limit cutoff. The Lyman limit absorption is not shifted into
the $V$ filter until redshifts above $z\simeq 4.5$.
The idea of using the partial Ly$\alpha$ forest absorption to identify
galaxies at $z\sim 2$ is the foundation of the so-called ``BX" method \citep{adelberger2004}. 
\citetalias{ly2011} has previously used the $U$, $B$ and $V$ photometry in
SDF to identify these BX galaxies. 

\begin{figure}
\begin{center}
\includegraphics[width=0.48\textwidth]{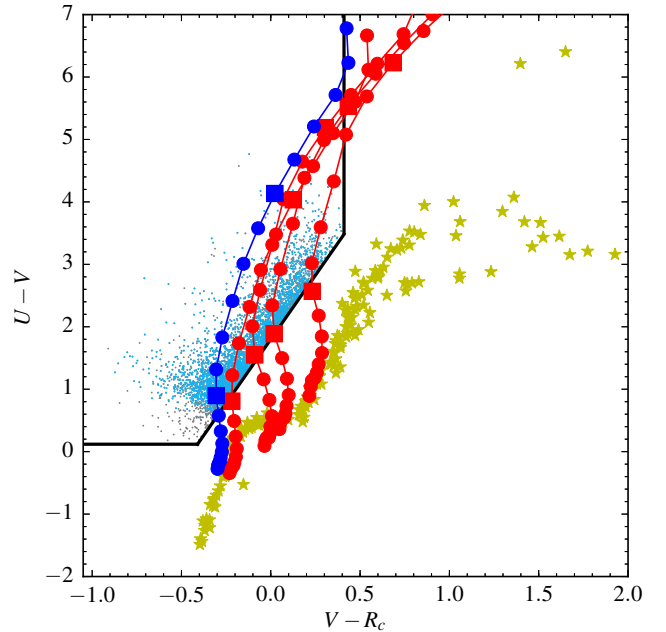}
\end{center}
\caption{%
$U-V$ vs. $V-R_c$ diagram for the selected LBGs,
shown as small dots. Cyan indicates a lower limit for $U$ (and thus $U-V$) whereas gray indicates
an object detected at $>1\sigma$ in $U$. 
The yellow stars mark 175 \citet{gunn1983} stars.
The thick black lines indicate the boundary that we adopt for
the selection of $z\sim 3$ LBGs.
Solid lines show example tracks for model spectra of young star-forming galaxies.
The redshift range of the models is from $z=2-4$,
with the circles on the track marking intervals of $\Delta z = 0.1$ (increasing upward in the diagram). 
The large squares 
on each track mark redshifts of $z=2.9$ and $z=3.5$.
The red tracks show colors of model spectra of
the solely the stellar continuum, with reddening of $E(B-V)=0.3,0.1,0$ and 0, going from right to left. 
The three right-most tracks are averages of galaxy templates between ages $10^8 - 10^{8.5}$ years, all with solar metallicity. The fourth red track from the right shows a very blue stellar continuum -- an average of ages $10^6 - 10^{6.5}$years and metallicity of 1/5 solar. The blue track (left-most curve) shows this bluest continuum model with the addition of Ly$\alpha$ emission, $W_0 = 20${\AA}, boosting flux in the $V$-band. We thus believe that our criteria can effectively select faint galaxies with strong nebular emission.
In our fitting to average SEDs of LBGs described below, 
we confirm that these stellar population parameters cover
the same range as the actual galaxies, at least on average.
\label{fig:color}}
\end{figure}

In \reffig{color}, we show the distribution of detected objects
in the $U-V$ versus $V-R_c$ diagram,
as well as predicted positions of high-redshift galaxies (described below)
and foreground objects (lower-redshift galaxies and Galactic stars).
We assigned the 1$\sigma$ limiting-magnitude for objects with $U$ and/or $V$ magnitudes fainter
than this level. %, as was done in previous studies. 
%this is also the procedure followed by Steidel

We set the selection criteria for LBGs at $z\sim 3$ as:
\begin{mathletters}
\begin{eqnarray}
  && 23.0 \le R_c\le 27.8, \\
  && U-V \ge 0.12,\\
  && U-V \ge 4.1 \times (V-R_c)  + 1.8,\\
  && \text{ and } V-R_c \le 0.41.
\end{eqnarray}
\end{mathletters}
The selection boundaries in the $U-V$ versus $V-R_c$ diagram
defined by these color criteria
are outlined with the thick black line in \reffig{color}.
The small dots represent all objects we measured in SDF
brighter than $R_c = 27.8$. 
Gray symbols represent galaxies detected in the U band; 
the more numerous (70\%
of the sample) blue symbols represent LBGs with upper limits in the U band,
meaning that their true locations in the diagram could be 
higher than what is shown.

To include extremely faint LBGs with low photometric
SNRs,  we accepted galaxies with formal U-V colors down to 0.12 mag.
This is bluer than allowed by some previous searches
for ``U dropouts". But it turned out that this made hardly
any difference.  If instead we had set the U-V cut at
$ \geq$ 0.50 mag, we would only have lost 48 LBG candidates--less
than 1\% 
of our final sample.  This would not have significantly changed any of our
results.

In general, our color selection is deliberately conservative. 
We have excluded relatively red LBGs from our sample 
because we wanted to minimize the possible contamination
from foreground interlopers. We have thus aimed to
construct a $z=3$ LBG sample of high purity.  Although its 
completeness may be lower than some previous studies,
we are able to estimate and correct for this effect (described in 
Section \ref{subsec:lbg-comp}).
We also excluded candidates brighter than $R_c= 23.0$
because, at that high brightness level,
extremely rare, very luminous LBGs would
likely be overwhelmed by spurious contaminants.

The data were compared with the predicted colors of model galaxies that are
expected to span the range of properties found in LBGs, shown as the red circles.
These models were produced using the stellar synthesis
code of \citet{bruzual2003} (hereafter BC03), assuming a Salpeter IMF.
%extending from 0.1 to 100 solar masses. 
We considered both constant and exponentially decaying
SFR histories.
In our modeling of
the stellar populations of LBGs, we have
considered three metallicities, $Z = 0.004, 0.008$ and 0.02 (solar),
and ages of 0.01, 0.1, 0.5, and 1 Gyr.
We assumed reddenings of $E(B-V)$ = 0.0, 0.1, 0.2, 0.3, and 0.4 and
the \citet{calzetti2000} attenuation curve.
Although we have no particular reason to believe that this
law is more applicable to LBGs than other possibilities,
we adopt it to facilitate comparison with many previous studies
which also assume the \citet{calzetti2000} curve.
The absorption due to the intergalactic medium was applied
following the prescription by \citet{madau1995}.
The colors were calculated by convolving the
constructed model spectra with the response functions of the
Suprime-Cam and MOSAIC broadband filters.
These models were computed at redshift intervals
of $\Delta z= 0.1$. These values reproduce the average rest-frame ultraviolet-optical
SED of LBGs observed at $z\sim 3$ \citep[e.g.,][]{papovich2001,shapley2001,ly2011}.

The red lines in \reffig{color} show some of our \citetalias{bruzual2003} stellar synthesis model tracks.
Regardless of the stellar population, our selection box is
sensitive to LBGs having redshifts of $z\simeq 2.9-3.5$ (the squares on the model tracks bracket this range).
The yellow stars show observed colors of \citet{gunn1983} stars.
For a given $V-R_c$ color, the stars are always bluer in $U-V$,
because they do not have so steep a ``$U$-drop'' as our LBGs. This is not the case 
for the $\mathcal{U}_n\mathcal{G}\mathcal{R}$ selection criterion, which is affected by contamination from main-sequence stars \citepalias{ly2011}.
The total number of LBGs selected is 5161, with a raw detected surface density of  $\approx$6 arcmin$^{-2}$. 

A small fraction of our LBG candidates
(roughly 15\%) have clear U dropouts (U -- V $\sim 1$), 
while showing extremely blue V -- R colors (of -0.3 to -0.4).
Their V -- R colors appear to be 0.1 -- 0.2 mag bluer than
our most extreme young starburst model.  
Closer examination reveals that these blue objects are 
entirely confined to our faintest LBG candidates (R $\geq$ 27), of 
which they constitute the majority.
They are, by definition, required to have solid detections
in the V band, but with otherwise quite noisy photometry.
Thus, given their large photometric uncertainties,
their true V -- R colors are probably somewhat redder than
shown in the diagram, in many cases.  As discussed below, we have modeled these
photometric uncertainties and find that our selection
of LBGs fainter than R = 27.0 is, indeed, extremely incomplete.
However, even at these faint magnitudes, we subsequently find that 
the contamination fraction of non-LBG interlopers must be
small.  This is demonstrated by the extremely strong K-band
excess that the stack of our 1294 faintest LBG candidates show,
attributable to [OIII]5007 emission at z$\sim 3$, discussed in
Section \ref{sec:LBGprops}.

Beyond photometry errors, there is an additional factor that 
may cause the extremely blue V -- R colors in our faintest LBG
candidates.  As discussed in Section \ref{subsubsec:nebem},
the strong nebular emission we find in all faint LBGs is
also seen in substantial Ly$\alpha$ emission, which
can be strong enough to increase the observed broadband V flux.
The blue track (left-most curve) shows the model of the youngest
stellar population with the addition of Ly$\alpha$ emission, $W_0 = 20${\AA}.
This Ly$\alpha$-enhanced track matches the colors of our faintest, bluest
LBG candidates.

\subsection{Number Counts} \label{subsec:lbg-num}

Table \ref{table:veff} gives the distribution of $R_c$-band magnitudes along with surface density
counts of our 5161 LBGs.  
Our sample includes $\sim$2000 LBGs fainter than $R_c=27.0$.
This gives us a significantly better statistical measure of the faint end of
the LBG LF than previous studies, which 
covered smaller areas or were less sensitive.

The surface density counts of our LBG candidates are consistent with previous studies,
except that we have a lower surface density of the brightest LBGs in the SDF.
Given the low surface density of such rare luminous galaxies, this apparent
discrepancy could be the result of cosmic variance. 
Another possibility is that these previous surveys might have suffered some contamination 
from interlopers at the bright end \citet{ly2011}.

\begin{table}[!ht]
\begin{center}
\caption{Magnitude distribution, Number Counts and effective survey volumes of the LBGs.
\label{table:veff}}
%\small
\begin{tabular}{lccccc}
\hline
\hline
$R_c$ & $N_{\rm{LBG}}$& $\Sigma$ (num arcmin$^{-2}$)\footnotemark[1] & $V_{\rm{eff}}$ $[\rm{Mpc^3}]$  \\ \hline
 23.0 - 23.5& 11 & 0.025 & 1.32$\ \times \ 10^6$ \\
 23.5 - 24.0& 49 & 0.112 & 1.26$\ \times \ 10^6$  \\
 24.0 - 24.5& 146 & 0.333 & 1.19$\ \times \ 10^6$  \\
 24.5 - 25.0& 334 & 0.762 & 0.94$\ \times \ 10^6$  \\
 25.0 - 25.5& 482 &  1.100 & 7.08$\ \times \ 10^5$  \\
 25.5 - 26.0& 590 & 1.347 & 4.81$\ \times \ 10^5$  \\
 26.0 - 26.5& 757 & 1.728 & 3.40$\ \times \ 10^5$  \\
 26.5 - 27.0& 929 & 2.121 & 2.52$\ \times \ 10^5$  \\
 27.0 - 27.5& 1132 & 2.584 & 2.22$\ \times \ 10^5$  \\
 27.5 - 28.0& 731 & 1.669 & 1.57$\ \times \ 10^5$  \\ \hline
 Total &5161& 11.783 & &
 \end{tabular}
\end{center}
\footnotetext[1]{Surface density of LBGs in each $R_c$-band magnitude bin. } 
%\label{table:veff}
\end{table}

\subsection{Redshift Distribution Function} \label{subsec:lbg-comp}

The redshift distribution function of the LBG sample and its completeness were estimated
as a function of magnitude through Monte Carlo simulation.
We generated artificial LBGs over an apparent
magnitude range of $23.0\le z' \le 27.5$, in 
intervals of $\Delta m=0.5$, and over a redshift range of
$2 \le z\le 4$, in intervals of $\Delta z = 0.1$.
Model spectra of the artificial LBGs are constructed
using the \citetalias{bruzual2003} stellar population synthesis code, as described above.
%As model parameters, an age of 0.1 Gyr, a Salpeter initial mass
%function, and a star-formation timescale of 5 Gyr are adopted,
%and five values of reddening, $E(B-V) = 0.0$, $0.1$, $0.2$, $0.3$,
%and $0.4$, are applied using the
%dust extinction formula for starburst galaxies by
%\citet{calzetti2000}.
%Thus, 650 ($5\times 25\times 5$) objects are generated.

The median size of our LBG candidates is
FWHM $\approx 1.''1$. Their unresolved or marginally
resolved morphology is consistent with previous
expectations that the intrinsic half-light diameters of
LBGs at $ z \sim 3$ should be about 2 kpc, or only a
few tenths of an arcsecond \citep{shibuya2015}.
There is a significant tail of resolved LBGs--about
10\% of them have FWHM $> 2''$. 
We assumed that the shape of LBGs is Gaussian.
We assigned apparent sizes to the artificial LBGs in our Monte Carlo simulation
matching the size distribution of observed LBG candidates measured by SExtractor
% (the peak FWHM is $\simeq 1\farcs41$-$1\farcs52$).

The artificial LBGs were then distributed randomly on the original
images after adding Poisson noise appropriate to their magnitudes, and
object detection and photometry were performed in the same manner
as done for real objects.
We generated 30,000 artificial galaxies on the image, and repeated this five
times, to obtain statistically accurate values of completeness.
In the simulations, the completeness for a given apparent magnitude,
redshift, and $E(B-V)$ value
is defined as the ratio in number of the simulated LBGs
that are detected and also satisfy the selection criteria to all
the simulated objects with the given magnitude, redshift,
and $E(B-V)$ value.
We calculated the completeness of the LBG sample
by taking a weighted average of the completeness
for each of the five $E(B-V)$ values.
The weight is taken using the $E(B-V)$ distribution function of
$z\sim 4$ LBGs derived by
\citet{ouchi2004b} (the open histogram in the bottom panel of their Figure 20),
which was corrected for
incompleteness due to selection biases.

This Monte Carlo simulation ignores the possible systematic 
positive correlation 
of $E(B-V)$ with galaxy luminosity, which is seen in observations
by \citet{labbe2007}, \citet{bouwens2009}, and \citet{sawicki2012}.
Our simple approach is the same as used by \citet{yoshida2006}, \citet{bian2013},
and also, judging from the descriptions they give, by
\citet{reddy2009} and \citet{hathi2010}. 
We and other groups have not included a correlation between $E(B-V)$
and magnitude in our simulations, mainly because determining the
true $\it intrinsic$ 
correlation is  difficult, given the selection effects
operating, and any correlation has such a large cosmic scatter
that it does not describe many LBGs.
The effects of different reddening assumptions have been
considered. \citet{reddy2008} found that their
``results are also insensitive to small variations in the assumed $E(B-V)$ distribution 
as long as the range of $E(B-V)$ chosen reflects that expected for the galaxies."
Similarly, \citet{sawicki2006} concluded that
``the dependence of the LF on these assumed dust and age values is negligibly small at z$\sim$4 and 3 but becomes more significant at the lower redshifts."

As shown in \reffig{color}, the diagonal color cut we used brings in
LBGs of all reddenings at nearly the identical redshift, z=2.9, shown
by the red squares.  This results in a sharp redshift cut-on for our 
LBG sample. The high-redshift cut-off is, at least in principle, not
so sharp. Theoretically, the bluest galaxies could still fall within
the selection box up to a redshift of z$\leq 3.7$, whereas the reddest
galaxies begin to leave the box at z$> 3.3$. 
This is a common feature of U dropout selections:  toward the flux
limits of the surveys, redder galaxies have a somewhat smaller chance
of being included. Our simple Monte Carlo simulation selected galaxy reddenings
independent of their magnitude.  A more sophisticated simulation might have
included a tendency for brighter LBGs to be redder, presumably because of
higher dust content \citep{steidel2003}. This could reveal a weak bias
in favor of discovering bluer LBGs at lower redshifts. 
We, along with others using two-color selections, have not included
such a correlation into our completeness simulations.  
To the extent that this correlation
is not merely a selection artifact of magnitude-limited surveys, its form
is quite uncertain, with a great deal of intrinsic scatter. In any case,
the reality, shown in the figures, is that our selection finds very few 
LBGs--of any type--at z$\ge 3.5$. There is therefore little room for
a possible (highly uncertain) color bias to undermine our completeness estimates.

\begin{figure}[!h]
\begin{center}
 \includegraphics[width=0.48\textwidth]{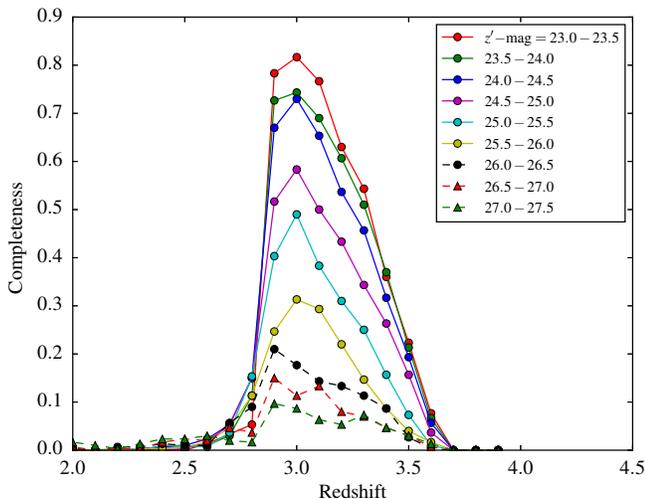}
\end{center}
\caption{
Redshift completeness functions $p(m,z)$ of the LBGs
with different apparent magnitudes
estimated from Monte Carlo simulations. The lines are color-coded
according to $z'$ magnitude. 
\label{fig:complete}}
\end{figure}

The resulting completeness, $p(m,z)$, is shown in \reffig{complete}.
Because we have been careful to make generous allowances for
the number of LBGs that could have been missed due to 
reddening, our {\it total} completeness is not extremely high,
dropping to 50\% at $R_c=25.0$ and 20\% at $R_c=26.0$.
This means that a large number of our fainter LBGs
are in a regime where our survey (and nearly all others)
is not very complete. The magnitude-weighted redshift distribution function
of our LBG sample was derived from $p(m,z)$
by averaging the magnitude-dependent completeness
weighted by the number of LBGs in each magnitude bin.
The average redshift and its standard deviation
are calculated to be $\langle z \rangle =3.3$ and $\sigma_z=0.3$.

\begin{figure}[!h]
\begin{center}
 \includegraphics[width=0.47\textwidth]{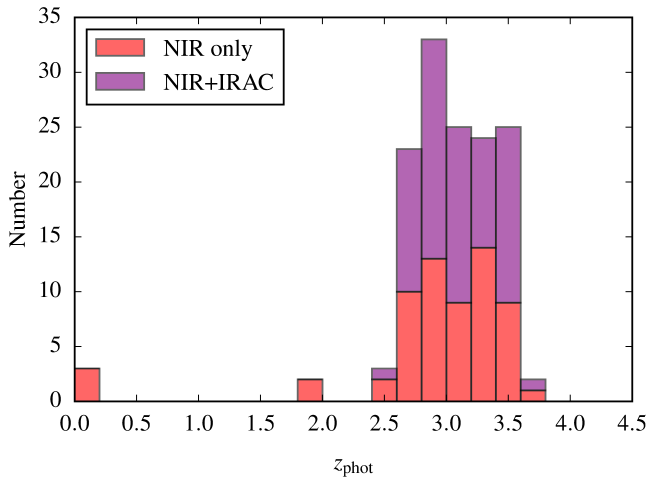}
\end{center}
\caption{%
Distribution of photometric redshifts for a subset of 140 LBGs detected in both the $H$ and
$K$ bands, 77 of which are also detected in IRAC bands 1 and 2 (purple bars in the histogram). The median 
brightness this NIR subset is $\langle R_c  \rangle \simeq 24.5$. 
\label{fig:zphot}}
\end{figure}

 A small minority (140) of the 5161 LBGs are individually detected ($>3\sigma$)
 in both of the WFCAM $H$ and $K$ band images (hereafter referred to as our NIR subsample). 
 Of these, slightly over half (77) are also detected in IRAC bands 1 and 2.
These are naturally the reddest LBGs in the parent sample and are expected to be the most massive (containing older stellar populations) or dust-obscured. 
So, although they are not representative of the full LBG population as a whole, they do have enough accurate
multi-band photometry over a wide wavelength range to allow us to measure their photometric redshift distribution. In particular, the $H$ and $K$ band photometry spans the Balmer break at these redshifts.
We input their optical-to-infrared SEDs into the Easy and Accurate Zphot from Yale \citep[EAZY;][]{eazy}. 
The resulting distribution of $z_{phot}$ for the NIR subsample is shown in Figure \ref{fig:zphot}. 
The NIR subsample has a median redshift  $\left< z_{phot} \right>_{med} = 3.03$ with $\sim 90\%$ of the LBG candidates
having photometric redshifts in the  range $z_{phot} = 2.5-3.5$. 
Thus, robust photometric redshifts derived for a subset of massive LBGs yield confidence in the effectiveness of our color selection 
(described above) to select $z\sim 3$ star-forming galaxies. We further discuss properties of the NIR subsample in Section \ref{sec:LBGprops}.

\subsection{Comparison with $\mathcal{U}_n-\mathcal{G}$ Dropouts in SDF} \label{subsec:ugdrops}
The same $UBVR_ci'z'$ photometry in SDF was previously used by \citetalias{ly2011}
to identify LBGs.  That study aimed to duplicate exactly
the $U$-dropout method used by Steidel and collaborators, which selects
objects with very red $\mathcal{U}_n-\mathcal{G}$ colors \citep[e.g.,][]{steidel1999}.
\citetalias{ly2011} approximated the $\mathcal{G}$ filter with a linear combination
of $B$ and $V$ photometry.  Because the $G$ band is about 500 {\AA}
bluer than the $V$ band used in this paper, their LBG sample starts at
$z=2.5$ rather than our $z=2.8$.  This results in their finding a 
higher surface density of bright $U$-dropouts than we did 
with our $U-V$ selection.
However, when we account for the different selection functions
and cosmic volumes surveyed, our resulting LF
is very similar to that of $\mathcal{U}_n-\mathcal{G}$ dropouts (see Section \ref{sec:lf} below).
%right, our wider selection function surveys larger cosmic volume per given
%area, but we correct for this in the final LF, which AGREES with Steidel
 In contrast to our selection down to $R_c = 27.8$, \citetalias{ly2011} cut their LBG selection at
relatively bright $U$-dropouts, having $R \leq 25.5$.
They also used a combination of photometric and image size information
to exclude stars, which removed about 10\% of the LBG
candidates at $R<24$, and a smaller, negligible fraction fainter than
that.  Overall, from cross-matching our $U-V$ dropout catalog with the $\mathcal{U}_n-\mathcal{G}$ dropout
catalog in \citetalias{ly2011}, we find only 651 sources in common (13\% of our
sample and 27\% of the \citetalias{ly2011} sample). This lack of overlap is due to
the effects mentioned above -- primarily
the differing color selections. Our sample occupies a
bluer, fainter locus on the $U-V$ vs. $V-R_c$ diagram than the $\mathcal{U}_n-\mathcal{G}$ dropouts
from \citetalias{ly2011}.  For comparison, we included the \citetalias{ly2011} sample in our analysis of
average UV-to-IR SEDs and inferred 
physical properties discussed in Section \ref{sec:LBGprops}. 

% \subsection{Comparison with GALEX NUV Dropouts} \label{subsec:galex}
% \citet{ly2009} used extremely deep imaging of the SDF with the 
% orbiting GALEX ultraviolet telescope to
% detect a sample of 8000 Near-ultraviolet (2500\AA)
% dropouts which are mainly Lyman break galaxies with
% redshifts of 1.5 to 3.
% This LBG sample goes down to a limiting B magnitude of 25.xx.
% Ly's simulations indicated that about 25\%
% or xxx00 of their LBGs should have redshifts above 2.7,
% and should thus also be U-band dropouts.  
% Restricting our U-dropout sample to galaxies bright
% enough to satisfy Ly et al's magnitude limit, we find a total
% of xxx Lyman break galaxies.  Thus our rate of U-dropouts
% in SDF is roughly consistent with that predicted by the Ly et al
% GALEX study. 
% only IFF chun will take the time to match his GALEX dropouts with our U-V dropouts xxx

%---------------------------------------------------------------------
\section{LBG Luminosity Function in the SDF} \label{sec:lf}

The LF for the LBG sample at $z\sim 3$ was
derived in the same manner as in \citet{steidel1999} and \citet{yoshida2006},
using the completeness estimates described in Section \ref{subsec:lbg-comp}.  
To estimate contamination, we took the fraction of low-redshift interlopers in the LBG
sample from the Monte Carlo simulation performed in \citet{yoshida2008}.
With an adopted boundary redshift of $z_0=2.9$ between interlopers and
LBGs, they find the fraction of low-redshift interlopers 
to be, at most, 6\% at any magnitude.
% below was originally excluded:
%We use objects in the Hubble Deep Field North (HDFN), for which
%best-fit spectra and photometric redshifts are given by
%\citet{furusawa2000}, as a template of the color, magnitude, and
%redshift distribution of foreground galaxies, and generate 929
%artificial objects which mimic the HDFN objects.
%The apparent sizes of the artificial objects are adjusted so that
%the size distribution recovered from the simulation is similar to
%that of the real objects in our catalogs.
%We distribute the artificial objects randomly on the original images
%after adding Poisson noise according to their magnitudes, and
%perform object detection and photometry in the same manner as employed for real objects.
%A sequence of these processes is repeated 100 times.
%In the simulation, the number of interlopers can be defined as the
%number of the simulated objects with low redshift ($z<z_0$) which are
%detected and also satisfy the selection criteria for LBGs.
%The number of interlopers expected in the LBG sample can then be
%calculated by multiplying the raw number by a scaling factor
%which corresponds to the ratio of the area of our field (876 arcmin$^2$) to
%the area of the HDFN multiplied by the repeated times (100 $\times $ 3.92 arcmin$^2$).
%\reffig{contam} shows the fraction
%of interlopers as a function of magnitude from \citet{yoshida2008}.
%and our contamination rate is expected to be very similar to theirs.

The effective volume is defined as:
\begin{equation}
V_{\rm{eff}}(m)=\int _{z_0}^{\infty}p(m,z)\frac{dV(z)}{dz}dz,
\end{equation}
where $p(m,z)$ is the completeness function described in Section \ref{subsec:lbg-comp}. We estimated this for our 876 arcmin$^2$ field using:
\begin{equation}
\frac{dV(z)}{dz}=\left(\frac{D_L}{1+z}\right)^2\times876\times\left(\frac{\pi}{10800}\right)^2,
\end{equation}
where $D_L$ is the luminosity distance. The quantity $dz dV/dz$ is the comoving volume per arcmin$^2$ at redshift $z$ (see \citet{steidel1999} for discussion). The effective volumes for each magnitude range and the number
of observed LBGs are given in Table \ref{table:veff}. The absolute magnitudes of the LBGs were determined assuming a flat UV continuum (i.e. $\lambda f_\lambda$ [erg s$^{-1}$cm$^{-2}$ {\AA}]$^{-1}$ is constant), such that the apparent UV magnitudes of the LBGs, $m_{\mathrm{UV}}$, are simply their $R_c$-band magnitudes (corresponding to rest-frame $\lambda \sim 1600$ {\AA}). Then:
\begin{equation} 
M_{\mathrm{UV}} = m_{\mathrm{UV}} - 5\log_{10}\left(\frac{D_L}{10\text{ pc}}\right) + 2.5 \log_{10}(1+z),
\end{equation}
with an assumed redshift of $z=3$. Note that the $K$-correction term is excluded, as it is nearly 0 \citep{sawicki2006}.

\begin{figure*}[ht]
\begin{center}
 \includegraphics[scale=0.65]{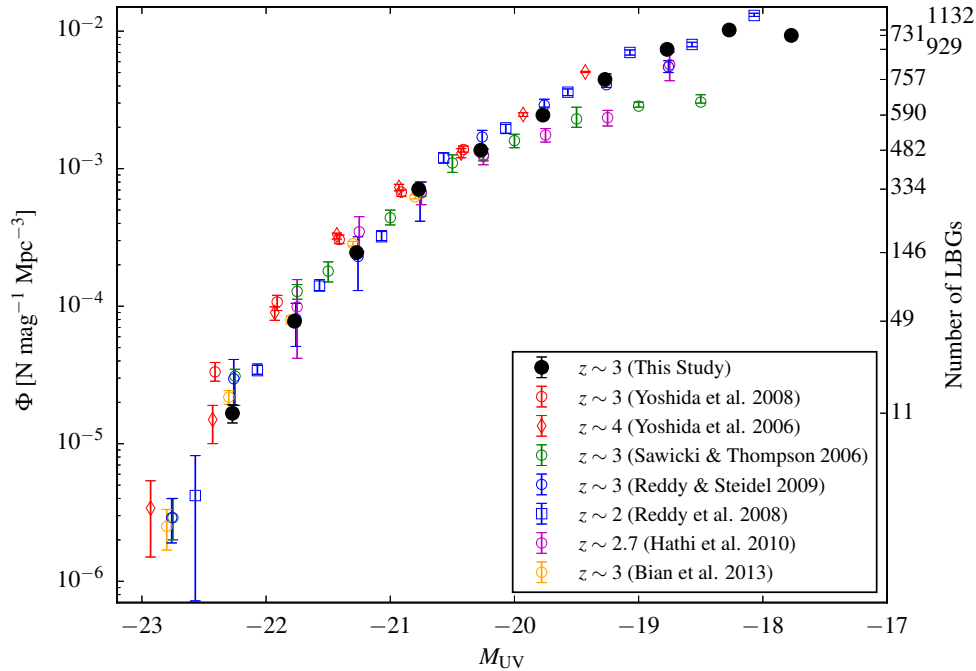}
\end{center}
\caption{%
UV luminosity function for our LBGs, compared to other
LBG samples. Circles correspond to measurements at $z\sim3$, diamonds to $z\sim 4$, and squares to $z\sim 2$. The right axis gives the number of LBGs in each magnitude bin for our sample. 
\label{fig:LF}}
\end{figure*}

\reffig{LF} shows our LBG LF in comparison to those from other
studies of galaxies at $z=3$
\citep{steidel1999,sawicki2006,yoshida2008,reddy2009,hathi2010,bian2013}.
More determinations of the LBG LF have been made by \citet{rafelski2009} and \citet{vanderBurg2010}. 
Our LF mostly agrees well with those in the literature,
particularly when one considers the different sample selections. The LFs from \citet{sawicki2006}
and  \citet{hathi2010} appear to be significantly lower at the faint end,
perhaps due to insufficient corrections for incompleteness.

Interestingly, the SDF $z\sim 3$ LBG LF we observe shows only a very gradual downward
curvature.  Therefore, there is not a very clearly defined ``knee''
at which it turns over, as in a Schechter function.
In fact, a double power law would likely give
a better fit to the data, with a relatively small flattening
of slope at low luminosities.
Nonetheless, for comparison with previous 
studies, we fit our LF with the standard Schechter function,
described by:
\begin{eqnarray}
\phi(M_{\rm{UV}})dM_{\rm{UV}} &&= \phi^*\left( \frac{\rm{ln} 10}{2.5}\right)\left(10^{-0.4(M_{\rm{UV}}-M^{*}_{\rm{UV}})}\right)^{\alpha+1} \\ && \times \exp \left(-10^{-0.4(M_{\rm{UV}}-M^{*}_{\rm{UV}})}\right)dM_{\rm{UV}},\nonumber
\end{eqnarray}
where $M_{\mathrm{UV}}^*$ is the characteristic absolute magnitude, 
$\phi^*$ is the normalization, and $\alpha$ is the faint-end slope. 
The best-fitting Schechter function parameters (along with those derived for other LBG studies) are given in Table
\ref{table:schechter}. Indeed we derive a steep faint-end slope of $\alpha=-1.78 \pm 0.05$ indicative of the high abundance of faint, low-mass systems in the universe. We note that \citet{bian2013} derives an even steeper faint-end slope of $\alpha=-1.94 \pm 0.10$; however, their sample is much shallower than ours, only reaching $M_{\mathrm{UV}} = -20.8$.

\begin{table*}[!ht]
\begin{center}
\caption{Schechter parameters of UV LF at $z\sim3$.\label{table:schechter}}
\small
\begin{threeparttable}
\begin{tabular}{cccccc}
\hline
\hline
Data & $M^*_{\rm{UV}}$ & $\log_{10}(\phi^*/\mathrm{Mpc}^{-3})$ & $\alpha^{\rm a}$ & $m_{{\rm lim}}$ &FoV [$\rm{arcmin^2}$] \\ \hline
This study &-20.86$\pm$0.11 & -2.73$^{+0.07}_{-0.08}$ & -1.78$\pm$0.05 &
$R_c<$27.8 & 876 \\
\citet{reddy2009}& -20.97$\pm$0.14 & -2.77$^{+0.12}_{-0.16}$
&-1.73$\pm$0.13 &${\cal R}<$26.5 & 3261 \\
\citet{sawicki2006} & -20.90$^{+0.22}_{-0.14}$ &
-2.77$^{+0.13}_{-0.09}$ & -1.43$^{+0.17}_{-0.09}$ &${\cal R}<$ 27.0 &
169 \\ 
\citet{bian2013} & -21.11$\pm 0.08$ &  -2.97$^{+0.12}_{-0.16}$ & -1.94$\pm 0.10$ &  $R<25.0$ & 470  \\ 
\hline
\end{tabular}
\begin{tablenotes}[para,flushleft]
   {\footnotesize ${}^{\rm a}$Our slope uncertainty does not include the possible bias toward including more very faint galaxies
if they have systematically lower reddenings than brighter LBGs. Such an effect could reduce the faint-end
LF slope by an amount comparable to the random uncertainty.}
\end{tablenotes}
\end{threeparttable}
\end{center}
\end{table*}

%---------------------------------------------------------------------
\section{Physical Properties of LBGs} \label{sec:LBGprops}
We inferred physical properties of the LBG sample from their UV-to-IR
SEDs. As described in Section \ref{subsec:lbg-comp}, a subset of 140 LBGs 
are detected in $H$ and $K$ bands with about half of these galaxies detected in IRAC bands 1 and 2 as well. 
However, the majority of our LBGs have relatively blue colors and are too faint to 
be detected in the IR. In order to study our sample as a whole, we obtained average detections of LBGs 
in the infrared by stacking on their optical positions in the long-wavelength images. The large number of 
LBGs in our sample allows us to probe LBG properties such as stellar mass down to very faint magnitudes. 

\subsection{LBG Stacking} \label{subsec:stacking}

To stack LBGs in the infrared, we divided our sample into
bins of $i'$-magnitude, a proxy of SFR at $z\sim 3$ (probing rest-frame $\sim$1900 {\AA}). 
Specifically, we stacked our sample of LBGs in bins of $i'=23.5-24.5$,
$24.5-25.5$, $25.5-26.5$ and $26.5-27.5$. These magnitude bins contain 249, 806,
1256, and 1588 LBGs, respectively. The $\sim 1200$ LBGs with $i'>27.5$ did not yield stacked 
detections, and are thus excluded from the following analysis. 
An image of a representative galaxy in each magnitude bin was formed by finding the median of each pixel in
stacks of roughly $30''\times 30''$ cut-outs, centered around the optical position of each LBG in the bin. Here, 
we use the median of pixels rather than the mean, as the latter is significantly more prone to outliers. 

\begin{figure*}[!h]
    \centering
    \begin{subfigure}
        \centering
        \includegraphics[width=0.45\textwidth]{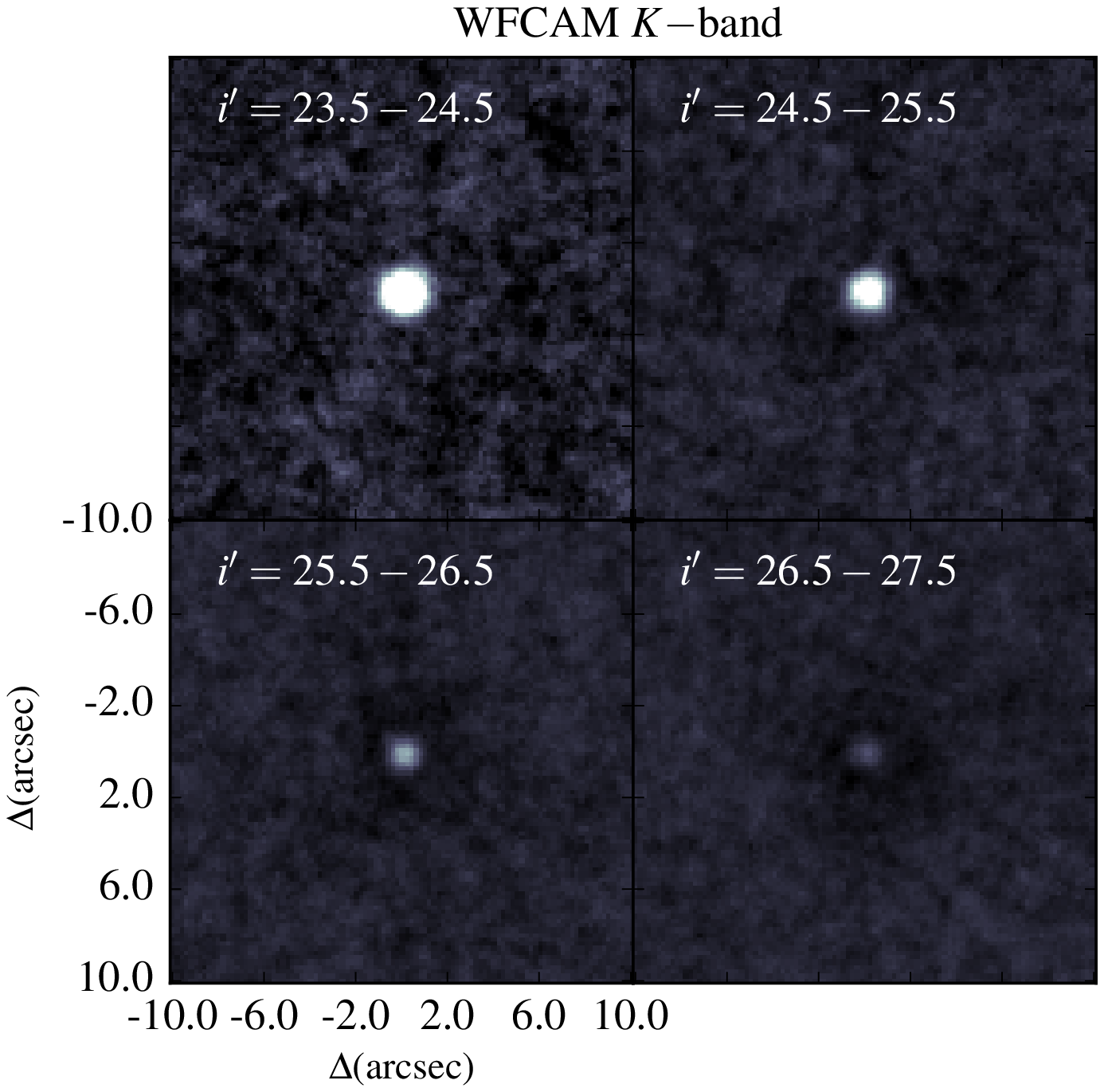}
    \end{subfigure}%
    ~ 
    \begin{subfigure}
        \centering
        \includegraphics[width=0.45\textwidth]{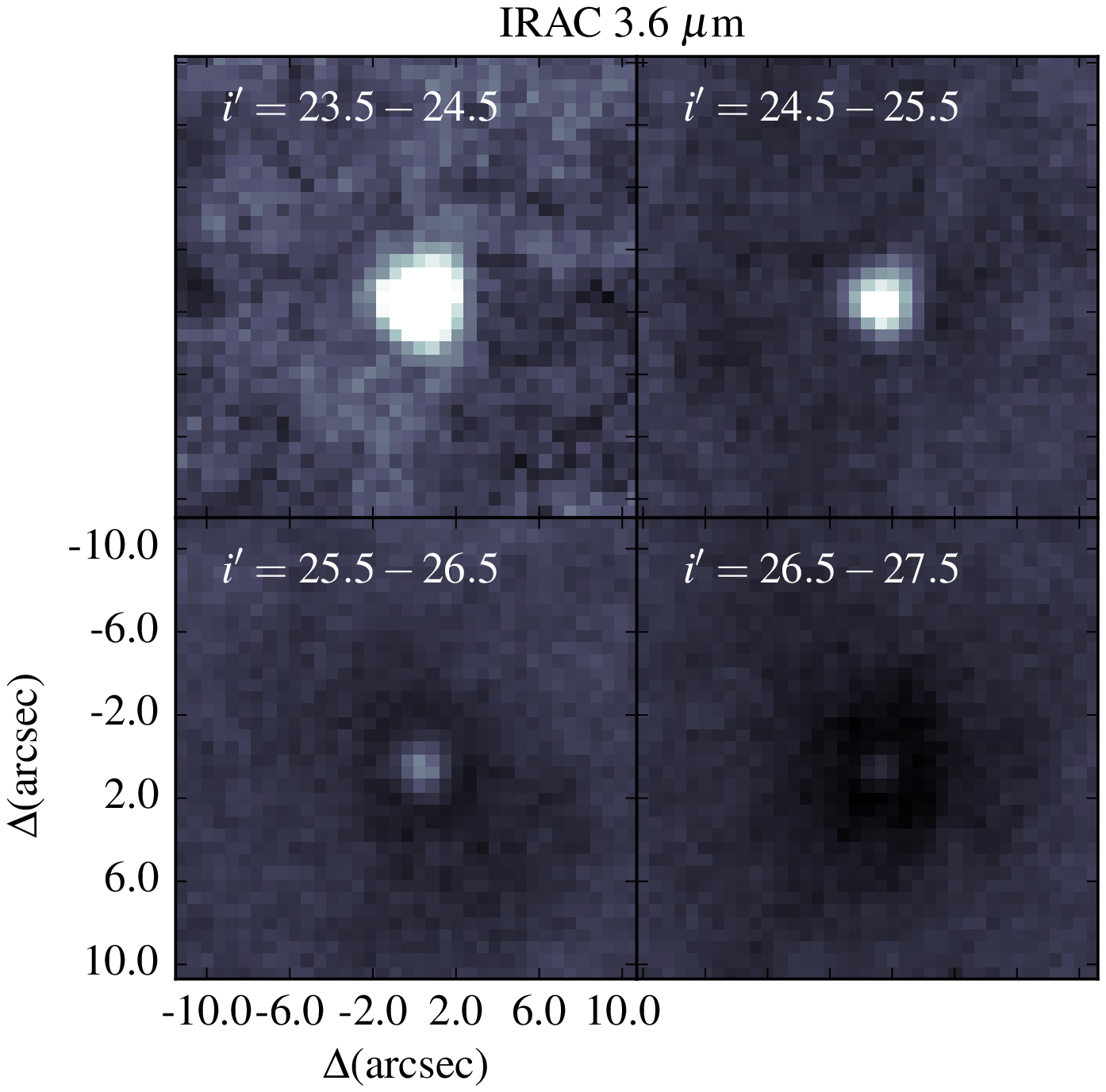}
    \end{subfigure}
    \caption{Stacks of LBGs binned by $i'$ magnitude. The left and
      right sets of panels show stacks of WFCAM $K$-band and IRAC 3.6
      $\mu$m data, respectively. Apparent in the fainter stacks is a depression in the background surrounding
      the stacked LBG. This defect is described in Section \ref{subsubsec:stackcorr}.    \label{fig:stackedims}}
\end{figure*}

Stacking is performed in the WFCAM $J$, $H$, and
$K$ bands, along with the NEWFIRM $H$ band, and all four IRAC
mosaics ([3.6], [4.5], [5.8], and [8.0]). Stacked fluxes were obtained from 
aperture photometry on the stacked images. We applied aperture corrections
 that were determined for
the WFCAM/NEWFIRM images of the SDF by identifying bright/isolated point sources
and computing the median curve of growth. For the IRAC mosaics, we used
the aperture corrections given in the IRAC Instrument Handbook\footnote{The corrections for 
a 4.8$''$-diameter aperture are -0.21, -0.22, 
-0.33, and -0.48 mag in the [3.6], [4.5], [5.8], and [8.0] mosaics, respectively}. 
As a check, 
we calculated our own aperture corrections and found that
they are within 0.1 mag of those quoted in the
handbook. Examples of stacked images of our LBGs in the
$K$-band and IRAC 3.6 $\mu$m mosaics are shown in Figure
\ref{fig:stackedims}. 
%We also perform stacking in the {\it
 % Spitzer}/MIPS 24 $\mu$m map and obtain upper limits (with no stacked detection of LBGs in
%these wavelengths). 

For each image, we computed the statistical significance of stacked flux as follows. 
First, we generated a set of $N$ random image coordinates (in the same regions used for stacking), 
formed a median-stacked image for these $N$ random coordinates, and found the aperture flux
 (with the same aperture size used to
measure our LBG stacked flux). We then repeated this process for 1000 iterations to form a
distribution of aperture flux from stacks of random image
positions. The width of this distribution (which is Gaussian) gives the 1$\sigma$ flux level 
corresponding to aperture photometry for stacks of $N$ coordinates on a given image of the SDF. 
Thus, we derived curves of 1$\sigma$ flux
vs. number of coordinates in a stack $N$ for each image. As
expected, we find that every curve follows a $1/\sqrt{N}$ law. For each band, the 3$\sigma$ limiting-magnitudes from stacking
$N=1000$ sources, after aperture correction, are: 
$J=27.44$, $H(\mathrm{WFCAM}) = 26.86 $, $H(\mathrm{NEWFIRM})= 26.74$, $K= 27.12$, 
$[3.6]= 26.34$, $[4.5]=26.35$, $[5.8]=24.82$, and $[8.0]=24.85$. 

\subsubsection{Faint-stack Defect and Corrections} \label{subsubsec:stackcorr}

In the stacks of fainter bins, such as in those displayed in Figure \ref{fig:stackedims}, we
encounter perhaps the largest source of uncertainty in stacked flux: a depression in
the background within the immediate vicinity ($\lesssim 10''$) of the detected
source. To understand
the cause of this defect and determine an appropriate method to
correct for it, we performed the following analysis. First, we used publicly
available SDF SExtractor detection
catalogs\footnote{\href{http://soaps.nao.ac.jp/SDF/v1/index.html}{http://soaps.nao.ac.jp/SDF/v1/index.html}}
and stacked on random source positions listed in the catalog, binning by $i'$
magnitude (down to total mag of $i'=29$). The defect remains present in
these test stacks of a very heterogeneous sample, so we rule out the defect
as being a result of our LBG selection technique. We find that the
defect is present in the faint $i'$-mag stacks of SDF images in {\it all} wavebands, including the
publicly available $BVR_ci'z'$ images. The deficit gets more severe with fainter source 
magnitude, and is perhaps most significant in the IRAC [3.6] and [4.5] bands. 

\begin{figure*}[hb]
\begin{center}
 \includegraphics[scale=0.6]{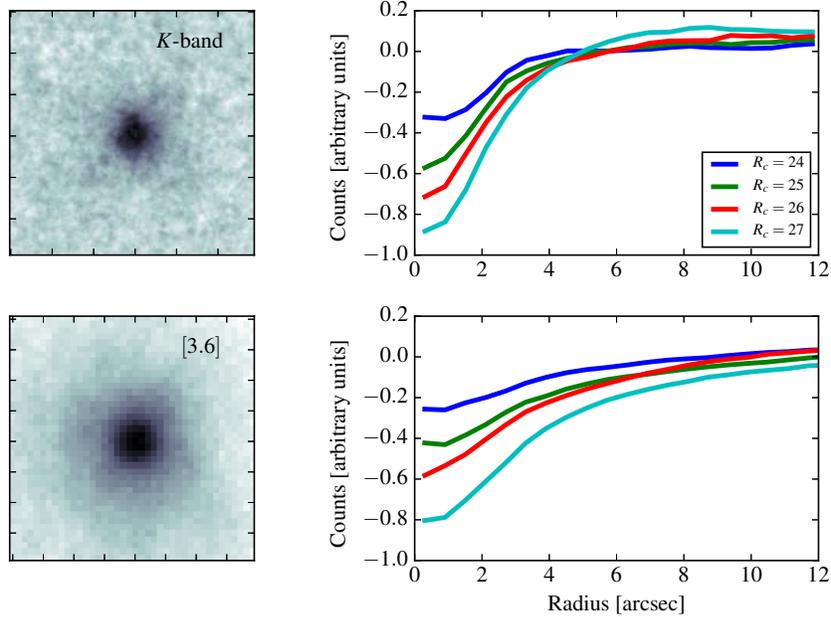}
\end{center}
\caption{Illustration of the background deficiency that occurs for stacks on positions of faint objects (see Section \ref{subsubsec:stackcorr}). Fifty thousand fake objects were added to the Suprime-Cam $R_c$-band image at random coordinates, and SExtractor was used to obtain a list of coordinates of {\it recovered} objects which, due to the detection algorithm, is a catalog of coordinates with relatively low background compared to random positions in the image. The left panels show stacked $K$-band and [3.6] images generated with the recovered coordinates out of the 50,000 identical $R_c$-mag$=26$ objects that were added to the detection image (but not added to the images used for stacking). The right panels show radial profiles of such images, and for the cases of the objects having magnitude of $R_c=24$, 25, and 27. The deficiency in background increases when attempting to stack on fainter objects in the detection catalog. 
\label{fig:stackdefect} }
\end{figure*}

From this test, we posit that the defect is caused by background counts around faint sources in
the detection catalogs being systematically lower than the background counts
around brighter sources. In other words, faint objects tend to enter these catalogs when they
are found in regions of the sky
with relatively low surface density of faint, blended objects (i.e. low confusion). This bias is due to SExtractor requiring a lower local
background level (higher S/N) in order to include a faint source as a
detection. To test this hypothesis, we performed a test similar to the calculation of the completeness of our sample.  
We created 50,000 fake galaxies (Gaussian-shaped) with chosen total magnitude and size, and added them to random positions 
in the Suprime-Cam $R_c$-band image (the detection image), cataloging their random positions. 
Next, we ran SExtractor on the new detection image (with all the synthetic galaxies added), with parameters identical to 
those used for LBG detection, and recover the coordinates 
of all {\it detected} synthetic objects.\footnote{The recovery rate is roughly 60\%, down to 50\% for synthetic objects 
with $R_c=25$ and $R_c=26$, respectively.}  According to our hypothesis, these coordinates should be positions with 
relatively lower background for fainter objects in the detection image. 
Finally, we stacked on those positions in the IR and obtained a ``hole" representing the pure background deficiency 
(because the synthetic galaxies were added only to the detection image). 
This effect is shown in Figure \ref{fig:stackdefect}. 
Stacking on all fake object input positions (which are just random image coordinates) results in a uniform stack with no hole apparent. 
As expected, the amplitude (depth) of the hole gets larger for stacks on recovered positions of fainter synthetic objects. 
Furthermore, we find that the profile gets deeper as the fake galaxies added are increased in size. 

In summary, we confirm that faint objects in SExtractor detection catalogs are located in regions of lower local background 
(due to a lack of faint, blended sources) than the brighter sources in the catalog. 
This effect is also likely present in catalogs generated with other detection algorithms. 
Thus, in any study utilizing such catalogs for stacking analysis of faint objects, the stacked flux of these objects will be 
systematically underestimated if this defect is not accounted for. We also note that low-resolution images that are
deep enough to begin to approach the confusion limit are particularly affected by this bias. 

\begin{figure*}[!ht]
\begin{center}
 \includegraphics[scale=0.6]{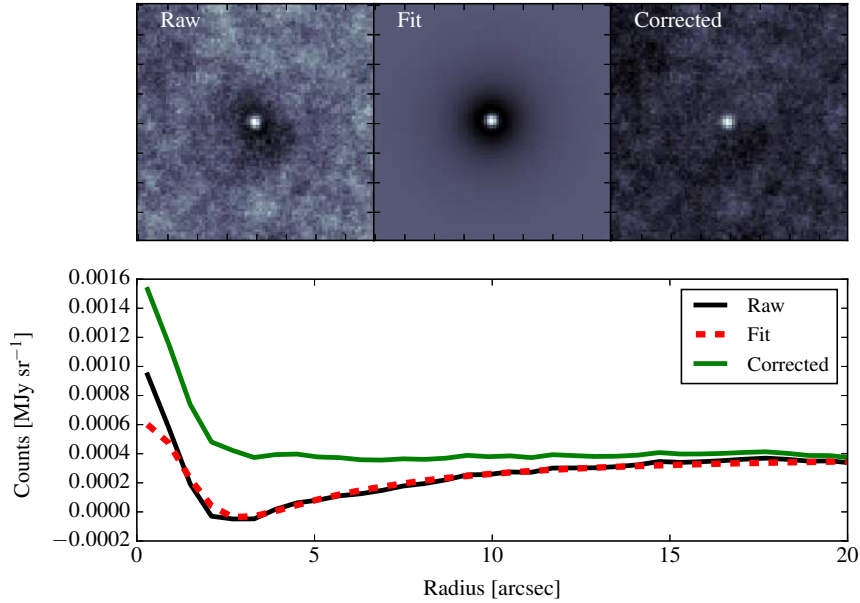}
\end{center}
\caption{Example of the procedure to correct faint stacked images for the 
background deficiency defect, applied here to the IRAC [3.6]
$\langle i' \rangle =26$ stack. The top panels show images, from left to right, of the raw
(uncorrected) image, the model (consisting of a positive Gaussian emission and
negative Lorentzian), and the corrected image with the negative
component subtracted out. The bottom panel displays the radial
profiles corresponding to the top images. 
\label{fig:stackcorr} }
\end{figure*}

This systematic loss of faint galaxies in the near vicinity of other galaxies was already corrected for in our LF calculations in Section \ref{sec:lf}, using our detailed simulations of incompleteness. However, corrections were still required for our stacked images in order to obtain accurate photometric fluxes. 

We performed a simple least-squares fit to each image with a two-component model consisting of a positive Gaussian function for the stacked emission and a negative Lorentzian component for the background depression. We fixed both components to be circularly symmetric with centroids on the center of the image. After the fit was performed, the corrected image was generated by subtracting the negative Lorentzian
component off of the original stacked image. An
example of this correction procedure is shown in Figure
\ref{fig:stackcorr} for the IRAC 3.6 $\mu$m stack in the $\langle i' \rangle =26$ bin, showing the raw image, the model, and the
corrected image along with corresponding radial profiles. Although the fit is good
for all stacked images, this correction is considered the primary source of uncertainty in
measuring the stacked flux of faint bins. 

%\newpage

\subsection{LBG Spectral Energy Distributions} \label{subsec:seds}

\begin{figure*}
\centering
 \includegraphics[scale=0.65]{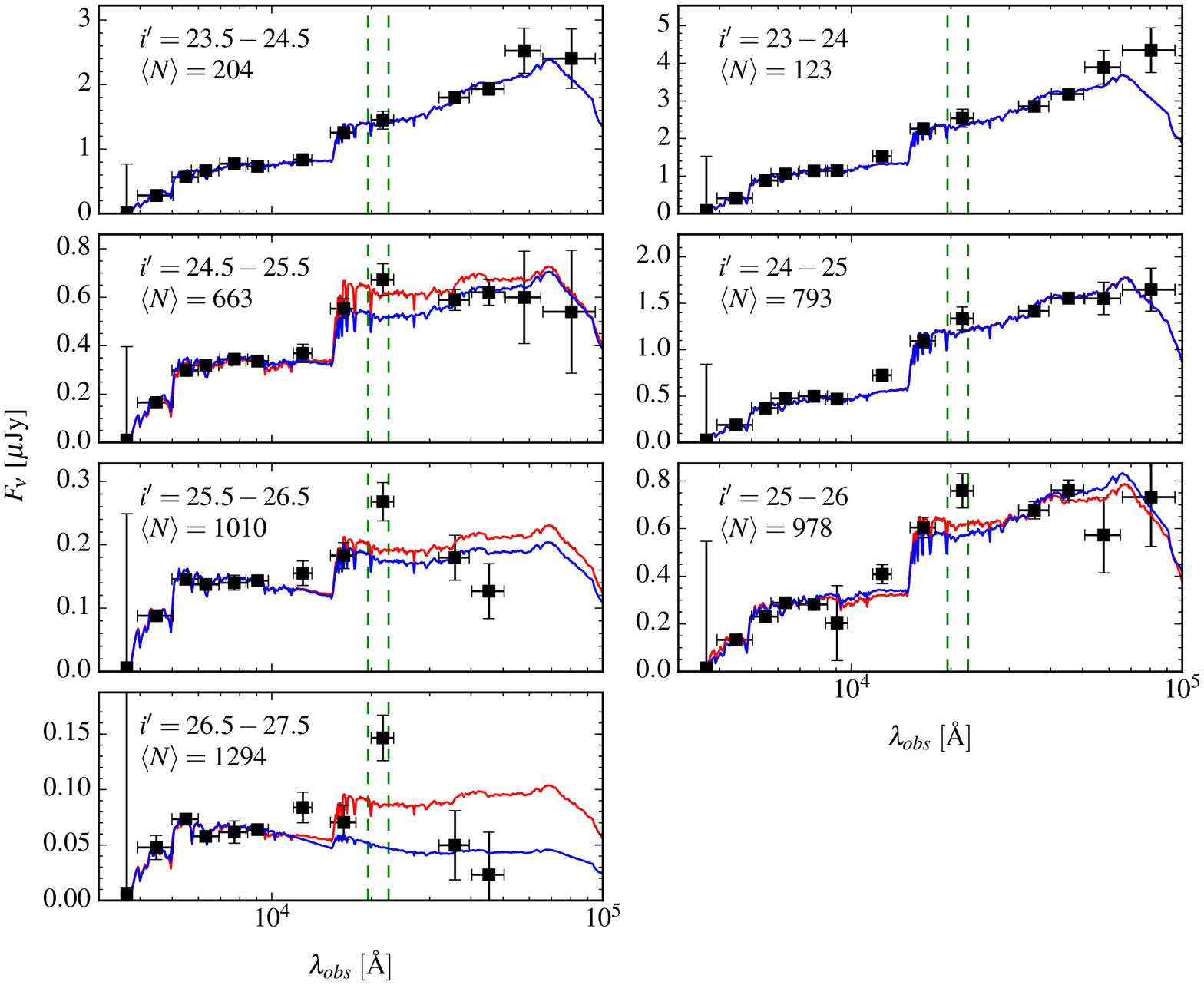}
  \caption{ Left: SEDs of the stacked $U-V$ dropout LBGs (binned by $i'$-magnitude) presented in this study. 
  Right: Stacked SEDs of $\mathcal{U}_n-\mathcal{G}$ dropout LBGs studied in \citet{ly2011}. In both
  panels, red curves show the best-fit stellar continuum model with the $K$-band data
  included, while blue curves show the best-fit model with $K$-band
  data excluded from the fit. Dashed lines bracket the range at which the [OIII]$\lambda\lambda$
  4959,5007 emission line doublet falls for $z=2.9-3.5$. We interpret the observed excess in the $K$-band 
  as contamination by [OIII] and H$\beta$. 
\label{fig:seds}}
\end{figure*}

We compute representative LBG SEDs using the median of $UBVR_ci'z'$ photometry 
and photometry from the median stacks of $J$, $H$, $K$, [3.6], [4.5], [5.8],
and [8.0] images in each $i'$-mag bin. Prior to measuring fluxes from stacked images, we implemented
 corrections to all of the NIR through mid-infrared stacks in the $\langle i' \rangle =25$ bin and fainter, in order to account for the faint stack background deficiency (described in Section \ref{subsubsec:stackcorr}). Error bars on stacked fluxes are calculated from the random stack analysis described above. The reported stacked $H$-band fluxes are formed from a weighted mean of the flux from the WFCAM and NEWFIRM stacked images. The average SEDs for our $U-V$ selected LBGs are plotted in the left panel of Figure \ref{fig:seds}. For comparison, we additionally performed stacking to form SEDs of the sample of $\mathcal{U}_n-\mathcal{G}$ selected LBGs from \citetalias{ly2011}. As explained in Section \ref{subsec:ugdrops}, this sample is redder and includes a higher proportion of galaxies on the lower-redshift end of our redshift distribution
function. We stacked this galaxy sample in three bins: 
$i'=23-24$, $24-25$, and $> 25$ (the sample is selected with $R < 25.5$, yielding sources down to $i' \simeq 26$). 
The faint stack defect is only 
non-negligible in the faintest bin $\langle i' \rangle=25.5$; in this bin all reported stacked data were
corrected using the method described above. The average SEDs for these $\mathcal{U}_n-\mathcal{G}$ dropouts are shown in the right panel of Figure \ref{fig:seds}. We also form the stacked SED of our massive, NIR-detected subset of 140 LBGs.

Average LBG properties as a function of $i'$ magnitude are derived using the Fitting and Assessment of Synthetic Templates (FAST) code \citep{kriek2009} to fit stellar population synthesis templates to each stacked
LBG SED. We use the \citetalias{bruzual2003} stellar population template
library, with a Salpeter IMF, and assume the \citet{calzetti2000} dust extinction
curve. We fit an exponentially decaying star-formation
history of the form $\mathrm{SFR} \sim \exp{(-t/\tau)}$. The ranges of parameter values for fitting
are set to: age $t=10^6 -10^{10}$ years, star formation history $\tau = 10^7-10^9$ years (in steps of 0.5 dex), 
and extinction $A_V = 0-3$. Metallicity is fixed at solar $Z = 0.02$. 
For our LBG sample, we fix the redshift of the fit to $z=3.1$, which is the approximate
peak of the completeness distribution for each magnitude bin. For the 
\citetalias{ly2011} sample, we fix the redshift to the median photometric redshift
in each bin. 
The resulting best-fit age, $\tau$, $A_V$, stellar mass $M_*$, and SFR
corresponding to each average LBG are given in Table \ref{tab:sedfitres}. Uncertainties for these physical
quantities are determined by FAST, using Monte Carlo simulations that redistribute the photometric data according to their error bars, and include additional uncertainty based on the uncertainties in the stellar continuum models (see the appendix of \citet{kriek2009} for more information).
 
The SEDs show a $K$-band excess that is more
pronounced for fainter bins. We attribute this excess to contamination
from redshifted [O{\sc iii}]$\lambda\lambda$4959,5007+H$\beta$ nebular line emission. 
As such, we perform fits both including and excluding the $K$-band photometry.
The best-fit models are shown in
Figure \ref{fig:seds} with the data; the blue and red curves exclude and include the $K$-band data, respectively.
The quoted stellar parameters in Table \ref{tab:sedfitres} correspond to the blue curves. 
Note that these fits only consist of a stellar continuum; 
we discuss the implications of nebular emission in Section \ref{subsec:prop}.

\begin{table*}[!t]
\begin{center}
\caption{Stellar properties of average $z\sim 3$ LBGs, derived from stellar-continuum 
fits to the stacked SEDs. \label{tab:sedfitres}}
\small
\begin{tabular}{ccccccc}
\hline  
\hline
$\langle i'  \rangle$\footnotemark[1] & $N_{\mathrm{stack}}$\footnotemark[2] &  $\log_{10}(M_*/{\rm M}_\odot)$  & 
SFR $[{\rm M}_\odot {\rm yr}^{-1}]$  & Age (Myr)  & 
$\tau$ (Myr)  &  $A_V$ \\
\hline 
\multicolumn{7}{c}{NIR-detected $U-V$ dropouts} \\
\hline
24.3 & $113-140$ & $10.47 \pm 0.06$ & $45$  $(+3,-17)$ & $398$  $(+233,-194)$  & 316 & 0.6 \\
\hline
\multicolumn{7}{c}{ $U-V$ dropouts} \\
\hline
24.2 & $169-217$ & $10.09$  $(+0.16,-0.33)$ & $51$  $(+48,-49)$ & $251$  $(+456,-208)$ & 1000 & 0.5 \\
25.1 & $589-724$ & $9.58$  $(+0.00,-0.26)$ & $12$  $(+0,-11)$ & $251$  $(+37,-192)$ & 316 & 0.2 \\
26.1 & $878-1078$ & $9.02$  $(+0.07,-0.69)$ & $3$  $(+22,-1)$ & $251$  $(+165,-242)$ & 316 & 0.0 \\
26.9 & $1138-1390$ & $7.79$  $(+0.74,-0.10)$ & $4$   $(+12,-3)$ & $10$  $(+253,-6)$ & 10 & 0.3 \\
\hline
\multicolumn{7}{c}{  $\mathcal{U}_n-\mathcal{G}$ dropouts \citepalias{ly2011} } \\
\hline
23.8 & $104-130$ & $10.22$  $(+0.12,-0.10)$ & $27$  $(+38,-25)$ & $100$  $(+157,-45)$ & 31 & 0.4 \\
24.7 & $668-867$ & $10.10$  $(+0.00,-0.30)$ & $13$  $(+0,-12)$ & $251$  $+0,-188)$ & 100 & 0.3 \\
25.3 & $828-1048$ & $9.54$  $(+0.14,-0.44)$ & $6$  $(+139,-6)$  & $100$  $(+175,-90)$ & 31 & 0.1 \\
\hline
\end{tabular}
\end{center}
\footnotetext[1]{Median $i'$ mag in bin.}
\footnotetext[2]{Range in number of objects in each stack. The minimum number of objects are in the WFCAM $H$-band stacks due to its incomplete coverage of the SDF.}
\end{table*}

\subsection{Stellar Properties and Nebular Emission} \label{subsec:prop}

 \begin{figure*}[!ht]
\begin{center}
 \includegraphics[scale=0.65]{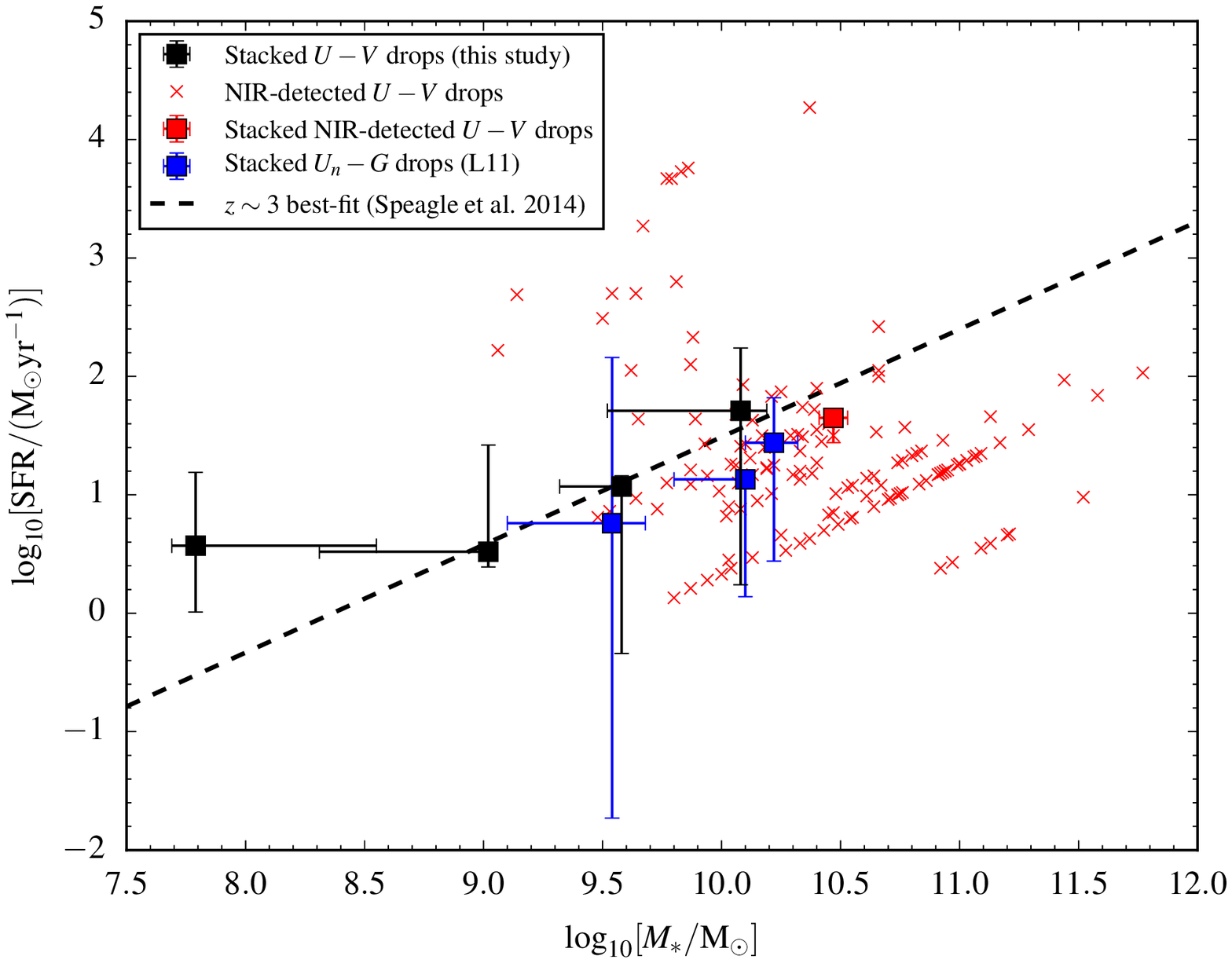}
\end{center}
\caption{Star formation rate versus stellar mass for the stacked LBGs in this study (black), the NIR-detected subset (individual as red X's and stacked as the red square), and of the stacks of $\mathcal{U}_n-\mathcal{G}$ dropouts (blue squares) presented in \citetalias{ly2011}. Also shown is an extrapolation of the best-fit line from \citet{speagle2014} to massive $z \simeq 3$ LBGs selected in \citet{magdis2010}, which we shift by a factor of 1.8 in mass in order to scale from their assumed Chabrier IMF to a Salpeter IMF. 
\label{fig:mainsequence} }
\end{figure*}

In general, we find that the properties derived for the LBGs studied here, given in Table \ref{tab:sedfitres}, are consistent with those derived in other studies of $z\sim 3$ star-forming galaxies. We compare our LBG sample to the ``star-forming main sequence" \citep[and references therein]{speagle2014} and plot their $M_* - \mathrm{SFR}$ correlation in Figure \ref{fig:mainsequence}. The dashed line in Figure \ref{fig:mainsequence} is a linear fit to a sample of massive (average stellar mass of $\simeq 5 \times 10^{10}$ $\mathrm{M}_\odot$) LBGs at $z\simeq 3$ studied in \citet{magdis2010}. Our average LBGs have stellar masses between $10^{7.8}$ and $10^{10.1}$ $\mathrm{M}_\odot$ and SFRs between 3 and 51 $\rm{M}_\odot$yr$^{-1}$. As expected, the stack of the NIR-detected sample is more massive, with $M_*\simeq 10^{10.5}$ $\rm{M}_\odot$ and $\mathrm{SFR}=45$ $M_\odot$ yr$^{-1}$.  The $\mathcal{U}_n-\mathcal{G}$ dropouts in \citetalias{ly2011} are found to be slightly more massive for a given $i'$ magnitude (which is also expected based on their redder colors). The average LBG in the faintest bin $\langle i' \rangle = 27$ has a best-fit model SED with low stellar mass $M_* \simeq 10^8$ M$_\odot$, very young age $t\simeq 10$ Myr, and high specific star formation rate (sSFR) of $10^{-7.2}$ yr$^{-1}$ or 60 Gyr$^{-1}$, placing it nearly $\sim 1$ dex above the fit to the \citet{magdis2010} sample.  Although the fitting parameters are highly uncertain, the properties are indicative of dwarf galaxies forming stars at a prolific rate.  
Their presence in large numbers suggests that the  ``star-forming sequence" in reality contains significant dispersion. The smaller the mass of the galaxy, the more stochastic its star formation episodes are likely to be, and the greater scatter this should produce in a $M_* - \mathrm{SFR}$ correlation. This insight is entirely consistent with what surveys at somewhat lower redshifts have found when they select galaxies by their line emission rather than broadband continuum \citep{atek2011,atek2014,ly2014,ly2015}.

 \subsubsection{ Nebular Emission in High sSFR LBGs} \label{subsubsec:nebem}
    Galaxies at the faint end of the LF, especially at high redshifts, have SEDs showing significant contributions from nebular emission \citep[e.g.,][]{yabe2009,atek2011,stark2013,deBarros2014}. Indeed, the observed average LBG SEDs in Figure \ref{fig:seds} contain some clear features that are not explained by the best-fit stellar continuum models. 
  
   Compared with filters on either side, the $K$-band magnitude is brighter by roughly 0.2, 0.4, and 0.9 mag for the $\langle i' \rangle =25$, 26, and 27 LBGs SEDs, respectively. The most likely explanation is that the stellar continuum in the $K$-band is boosted by a strong contribution from [OIII]$\lambda\lambda$4959,5007 emission at $z\simeq 3.1-3.8$ (with some contribution from $\rm{H}\beta$). Assuming this interpretation is correct, we estimate the equivalent width $\rm{EW}(\rm{[OIII]\lambda\lambda4959,5007}+\rm{H}\beta)$ for each average LBG. Specifically, we calculate the expected stellar continuum at the $K$-band wavelength, $m_{2.2\mu\mathrm{m},\mathrm{cont}}$, by convolving the best-fit stellar continuum models with the WFCAM $K$-band response curve. Knowing the $K$-band magnitude $m_K$ and the filter FWHM $\Delta K = 0.34$ $\mu$m, we then calculate the observed equivalent widths as:
  \begin{equation}
  \mathrm{EW}(\mathrm{[OIII]}+\mathrm{H}\beta) \simeq (1-10^{-0.4[m_K-m_{2.2\mu\mathrm{m},\mathrm{cont}}]})\cdot \Delta K.
  \end{equation}
Error bars on the equivalent widths are calculated by re-sampling the $H$, $K$, and [3.6] filter fluxes 
with random noise added based on their photometric errors. 
Assuming a redshift of $z=3.1$, the implied rest-frame equivalent widths of the $\langle i' \rangle = 25$, 26, and 27 stacked LBGs are $\rm{EW}_0(\rm{[OIII]}+\rm{H}\beta) \simeq 247 \pm 105$, $443 \pm 143$, and $1743 \pm 360$ {\AA}, respectively. 
  The $K$-band excess is also observed in the stacked SEDs for the $\mathcal{U}_n-\mathcal{G}$ dropouts from \citetalias{ly2011}, implying $\rm{EW}_0(\rm{[OIII]}+\rm{H}\beta) \simeq 79  \pm 86$ and $303 \pm 110$ {\AA} for the $\langle i' \rangle=24.5$ and 25.5 LBGs, respectively. 
  
In order to facilitate comparison with other samples, we remove the predicted contribution of H$\beta$ by assuming ratios for 
[OIII]$\lambda\lambda$4959,5007/H$\beta$. 
The [OIII]/H$\beta$ ratios were taken from mass-stacked spectra of galaxies in the WFC3 Infrared Spectroscopic Parallel survey (WISPS) in
\citet{henry2013} and \citet{dominguez2013}. 
Comparing the mean mass and resulting stacked [OIII]/H$\beta$ in these studies with the stellar mass of our stacked LBGs, 
we use ratios of [OIII]/H$\beta$$ = 3.0$, 4.5, and 5.0 for our three faintest bins of LBGs, and [OIII]/H$\beta$$ = 2.5$ 
and 3.0 for the two faintest bins of the \citetalias{ly2011} $\mathcal{U}_n-\mathcal{G}$ dropouts.
  
\begin{figure*}[!ht]
\begin{center}
 \includegraphics[scale=0.68]{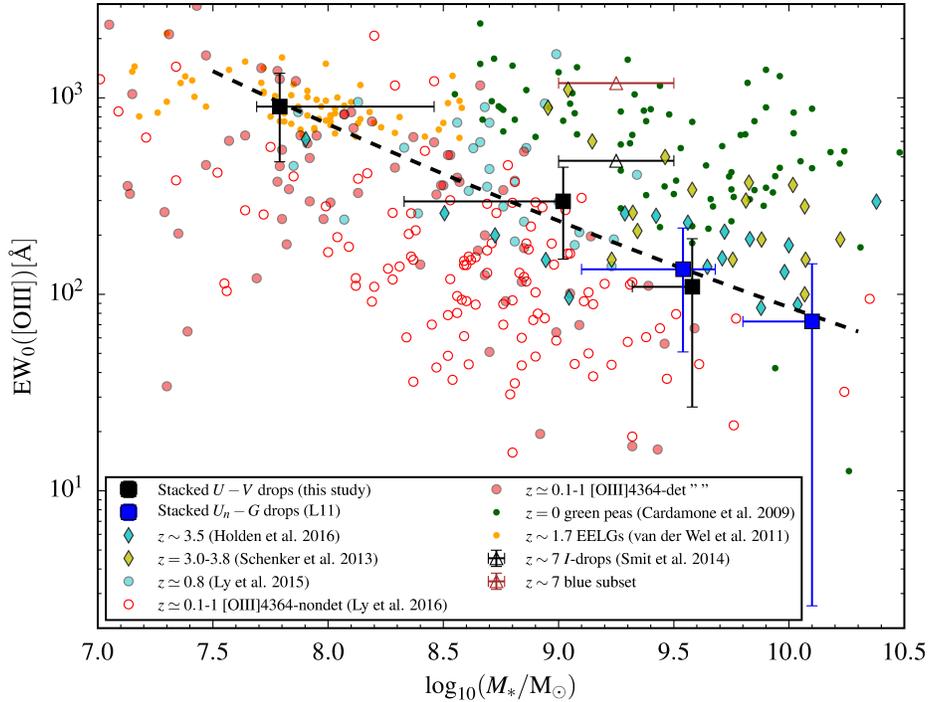}
\end{center}
\caption{ Rest-frame [OIII]$\lambda\lambda$4959,5007 equivalent width versus stellar mass, 
implied by the $K$-band flux excesses observed for the stacked LBGs in this study. 
To isolate the [OIII] doublet, we have assumed [OIII]/H$\beta$ ratios based on mass-stacked spectra of emission-line galaxies at lower redshift. 
The black dashed line represents a power law fitted to the stacked LBGs (both the $U-V$ dropouts here and stacks of $U-G_n$ 
dropouts selected in \citetalias{ly2011}). We overplot other samples of [OIII]-emitters (see Section \ref{subsubsec:nebem} for 
descriptions of each). For these samples, $\rm{EW}$s with only measurements for the ${\rm [OIII]}\lambda 5007$ line were 
increased by a factor of 1/3 to account for the 4959 {\AA} component. 
The \citet{smit2014} sample is overplotted assuming [OIII]/H$\beta$ = 4. 
\label{fig:WOIII} }
\end{figure*}

  Figure \ref{fig:WOIII} shows $\rm{EW}_0(\rm{[OIII]\lambda\lambda4959,5007})$ versus stellar mass for the LBGs studied here, compared with emission-line galaxy (ELG) samples at different redshifts. We find a steep dependence of $\rm{EW}_0(\rm{[OIII]})$ on stellar mass, which we quantify by fitting a simple power-law with amplitude $C$ and slope $p$:
  \begin{equation}
  \mathrm{EW}_0(\mathrm{[OIII]\lambda\lambda4959,5007}) = C [\log_{10}(M_*/\mathrm{M}_\odot)]^{-p}.
  \end{equation}
The least-squares fit, shown as the black dashed line in Figure \ref{fig:WOIII}, is performed for both the stacks of $U-V$ dropouts selected 
in this study and the stacks of $\mathcal{U}-{G}_n$ dropouts selected in \citetalias{ly2011}. 
The LBGs have a best-fit slope of $p = 10.2 \pm 1.5$. 
Although highly uncertain, the fit implies that the average $z\sim 3$ LBG with a stellar mass of $M_* = 10^{9}\mathrm{M}_\odot$ 
has a rest-frame [OIII]$\lambda\lambda$4959,5007 equivalent-width of $\simeq 340$ {\AA}, 
while a dwarf LBG with $M_* = 10^{8}\mathrm{M}_\odot$ has $\rm{EW}_0(\mathrm{[OIII]})\simeq 1130$ {\AA}. 
To emphasize how extraordinarily strong these [OIII] emission lines are in the faint LBGs, we note that this doublet is carrying several
percent of the entire luminosity of the galaxy.  This makes it one of the most distinctive observable features -- after the Lyman break -- in the entire galaxy spectrum.

Local analogs to the faint LBGs are the ``green pea" galaxies that were identified, in the SDSS, by their 
distinctive green color originating from strong [OIII]$\lambda\lambda$4959,5007 emission \citep{cardamone2009}. 
The SDSS green peas have masses in the range $\log_{10}(M_*/\mathrm{M}_\odot) = 8.5-10.5$ and typical sSFRs of $\simeq 4$ Gyr$^{-1}$, 
comparable to those derived for our average LBGs (excluding the faintest bin with extremely high implied sSFR). 
These local [OIII]-emitters are quite rare, with a rough surface density of $\sim 10^{-3}$ arcmin$^{-2}$. 
  
%\vspace{10mm}
    
    Less-massive ELGs at $z \lesssim 1$ are presented in \citet{ly2015} and \citet{ly2016}. In particular, \citet{ly2015} presents a sample of metal-poor, [OIII]$\lambda$4363-detected $z\simeq 0.8$ galaxy sample from the DEEP2 Galaxy Redshift Survey \citep{newman2013}. These galaxies have high sSFR near 10 Gyr$^{-1}$ and lie about 1 dex above the typical $M_*-\mathrm{SFR}$ correlation at $z\simeq 0.8$, similar to what is observed for the $\langle i' \rangle = 27$ stacked LBG in this study. This sample, shown on Figure \ref{fig:WOIII} as cyan circles, has a median $\mathrm{EW}_0\mathrm{([OIII]\lambda\lambda4959,5007}) \sim 400$ {\AA} and extends down to $M_* \simeq 10^8$ M$_\odot$. The galaxies at $z=0.1-1$ from \citet{ly2016}, selected by narrow-band emission in the SDF, are shown in Figure \ref{fig:WOIII} as red circles. The filled circles represent galaxies with detectable [OIII]$\lambda$4363 emission ($\simeq$36\% of their sample), while those without detection of this line are shown as open circles. Taken together, these ELGs have a median $\mathrm{EW}_0\mathrm{([OIII]\lambda\lambda4959,5007}) \simeq 160$ {\AA}.

  WISPS has produced larger samples of emission-line galaxies at $z\sim 1-2$. The properties of our fainter LBGs overlap with some of the more extreme emission-line galaxies in WISP. For example, \citet{atek2011} selected [OIII] and H$\alpha$ emitters with $\rm{EW} > 200$ {\AA} (rest-frame). The $\langle i' \rangle = 26$ average LBG in our study (with $\rm{ EW_0([OIII])}\simeq 360$ {\AA}) is roughly within the top half of the [OIII] EW distribution in \citet{atek2011}, while our faintest average LBG ($\langle i'\rangle = 27$) has $\rm{EW}_0(\mathrm{[OIII]})\simeq 1450$ {\AA}, which is in the top $\sim$15\%. A larger WISP sample, presented in \citet{atek2014}, shows that our average $\langle i' \rangle = 24, 25$, and $26$ LBGs have sSFRs in the top $\sim$40\% of these emission-line galaxies. Our faintest average LBG is near the top 5\% of their sSFR distribution. 

We also compare our LBGs to extreme emission-line galaxies (EELGs) at higher redshift from \citet{vanderWel2011} and \citet{maseda2014}. 
The galaxies in these studies were selected to have very large $\mathrm{EW}_0\mathrm{([OIII])}$, strong enough to produce detectable excess in the $J$ and $H$ bands for their samples at $z\sim 1.7$ and $z\sim 2.2$, respectively. These EELGs have typical masses of $\sim 10^8-10^9 M_\odot$ and high sSFRs ($\simeq 10$ Gyr$^{-1}$ for the galaxies in \citet{maseda2014} that are comparable to the best-fit sSFRs for our average LBGs. The $\langle i' \rangle =27 $ stacked LBG from our sample is within the locus of the \citet{vanderWel2011} EELGs in Figure \ref{fig:WOIII}. Interestingly, these galaxies have a space density nearly two orders of magnitude higher than the local green peas. 

The enhancement of $\mathrm{EW}_0\mathrm{([OIII])}$ is not only associated with low-mass dwarf galaxies 
with extreme starbursts,  it also appears to increase at higher redshifts across the board.
It has long been suspected that [OIII] emission is much stronger at higher redshifts ($z>3$),  than in star-forming galaxies at lower
redshifts.   This was the conclusion of the first narrow-band imaging search for [OIII]-line-emitting galaxies at $z=3.3$ (\citet{teplitz1999}).
Near-infrared spectroscopy further confirmed the unusual strength of [OIII] in Lyman break galaxies at $z\sim 3$ (\citet{teplitz2000}), and
gravitationally lensed galaxies at these redshifts hinted that [OIII] was even stronger in the less-luminous galaxies (\citet{teplitz2004}).

The next step forward came with multi-object spectroscopy, using MOSFIRE on Keck,
which showed that strong [OIII] emission is common amongst $z\sim 3-4$ LBGs. 
\citet{schenker2013} found strong $K$-band emission lines in 20/28 targeted LBGs at $z=3.0-3.8$ (13 of which had prior spectroscopic confirmation, and 15 of which were photometrically selected). 
These galaxies, which overlap the mass range of our sample,
have a median of $\mathrm{EW}_0(\mathrm{[OIII]4959,5007}) \simeq 280$ {\AA}, with a few reaching 
$\sim 1000$ {\AA} or higher.  %The $\langle i' \rangle = 25$ and 26 stacked LBGs in our study,
%would have [OIII] EWs on the low end of their distribution (see Figure \ref{fig:WOIII}). 
More recently, \citet{holden2016} present $K$-band spectra of 15 spectroscopically confirmed and 9 photometrically selected LBGs at $z=3.2-3.8$. Again, strong [OIII] emission is detected in 18/24 LBGs (15/15 and 3/9 for the spectroscopic and photometric samples, respectively), with a median $\mathrm{EW}_0(\mathrm{[OIII]}) \simeq 180$ {\AA}. As shown by Figure \ref{fig:WOIII}, our stacked LBGs are quite consistent with their EW distribution. Combined with our result, these studies suggest that the fraction of [OIII]-emitters in the LBG population is substantial at $3 \lesssim z \lesssim 4$. 
  
  Also shown in Figure \ref{fig:WOIII} (as black and brown open triangles) are the implied EWs from average SEDs of a sample of $I$-dropouts presented in \citet{smit2014}. These $z\sim 7$ galaxies (with masses of $10^9-10^{9.5}M_\odot$) show a very large excess in the IRAC [3.6] filter, which indicates rest-frame equivalent widths of $\rm{EW}_0(\rm{[OIII])}+\rm{H}\beta) > 637$ {\AA} for the average galaxy in the sample and $\rm{EW}_0(\rm{[OIII]}+\rm{H}\beta) \simeq 1582$ {\AA} for a subset of the bluest galaxies. Signatures of such strong [OIII]+H$\beta$ contamination are also reported for a handful of $z\sim 8$ galaxies with $M_*\simeq 10^9$ M$_\odot$ in \citet{labbe2013}, appearing as an excess in the IRAC [4.5] band. The implied equivalent width for their sample is $\mathrm{EW}_0(\rm{[OIII]}+\rm{H}\beta) = 670$ {\AA}.  Thus, we conclude that although very strong [OIII]-emitting galaxies are rare in the local universe, they become increasingly common from redshifts $1 \lesssim z \lesssim 2$, to $2 \lesssim z \lesssim 3$.  At higher redshifts, very strong [OIII] emitters may even become the norm.

 Aside from the $K$-band excess, we observe two more subtle features in the faint average LBG SEDs in Figure \ref{fig:seds} that are not explained by the best-fit stellar models. First, there is an excess in the $J$-band flux over the stellar continuum. An analogous feature has been found in stacked SEDs of $z \sim 4$ galaxies in \citet{gonzalez2012}, in the form an $H$-band excess. These authors attribute the excess to a bias due to the Balmer break falling into the $H$-band for the low-redshift tail of their sample, such that the stacked $H$-band flux is increased by flux redward of the break for the lower-redshift tail of their sample. However, our LBG selection is not sensitive to the low redshifts required for this bias to explain the observed stacked $J$-band excess. An alternate contribution to the observed $J$-band excess in our faint LBG SEDs might come from bound-free and free-free nebular continuum emission at rest-frame $\sim 3100$ {\AA}. We predict the possible contribution from nebular continuum emission at these wavelengths by assuming a typical $\text{[O\textsc{iii}]/H}\beta = 3$ to estimate the H$\beta$ flux from the [O{\sc iii}] equivalent widths. Using the tabulated H$\beta$ and continuum emission coefficients at an electron temperature of $T_e = 10,000$ K from \citet{osterbrock1989}, we determine that the nebular continuum can contribute $\sim 10$\% to the observed $J$-band flux. This would still leave about 20\% of the observed flux in the faintest bin unaccounted for. We thus conclude that the $J$-band excess is largely due to uncertainty in forming the stacked images, such as an over-correction for the faint-stack defect described in Section \ref{subsubsec:stackcorr}. Another uncertainty comes from our assumption of a simple stellar population to describe the starlight.

There is also weak evidence of Ly$\alpha$ line emission increasing $V$-band flux in the faintest LBGs ($\langle i' \rangle = 27$). Comparing the observed $V$ flux to the flux predicted by convolving the stellar continuum model with the $V$ filter transmission profile yields an excess of $\simeq 2\sigma$ significance, and implies a Ly$\alpha$ equivalent width of $\mathrm{EW}_0(\mathrm{Ly}\alpha) \simeq 30$ {\AA}.  If correct, this would mean that most of the LBGs in the faintest stack are Ly$\alpha$ emitters (LAEs). This would be consistent with the increasing LAE fraction found in the faintest LBGs at $3 < z < 7$  \citep{stark2010,stark2011,pentericci2011,schenker2012,ono2012,treu2012}. However, those studies generally do not include galaxies fainter than $M_{UV} = -18.75$, while the LBGs in our sample extend to lower luminosity. Nonetheless, if we extrapolate the result from \citet{stark2010} down to absolute magnitude $M_{UV} \simeq -18$, this would predict that $\sim 70$\% of LBGs at $3 < z < 6.3$ should be LAEs with $\mathrm{EW}_0(\mathrm{Ly}\alpha) > 50$ {\AA}. Our observed {\it average}\ 30 {\AA} equivalent width could be explained if about half of the faintest LBGs are strong LAEs with $\mathrm{EW}_0(\mathrm{Ly}\alpha) \gtrsim 50$ {\AA}. Alternatively, a few bright sources in the faint bin might have extremely strong Ly$\alpha$ emission, with the rest having weak or no emission (or perhaps even absorption). This would be consistent with what is observed for bright LBGs (\citep{shapley2003}, for example); however, we are in the regime of faint dwarf galaxies with little extinction from dust.

 \subsubsection{ [OIII] Line Ratios and ISM Conditions} \label{subsubsec:OIIIBalmer}

\begin{figure*}[!ht]
\begin{center}
 \includegraphics[scale=0.7]{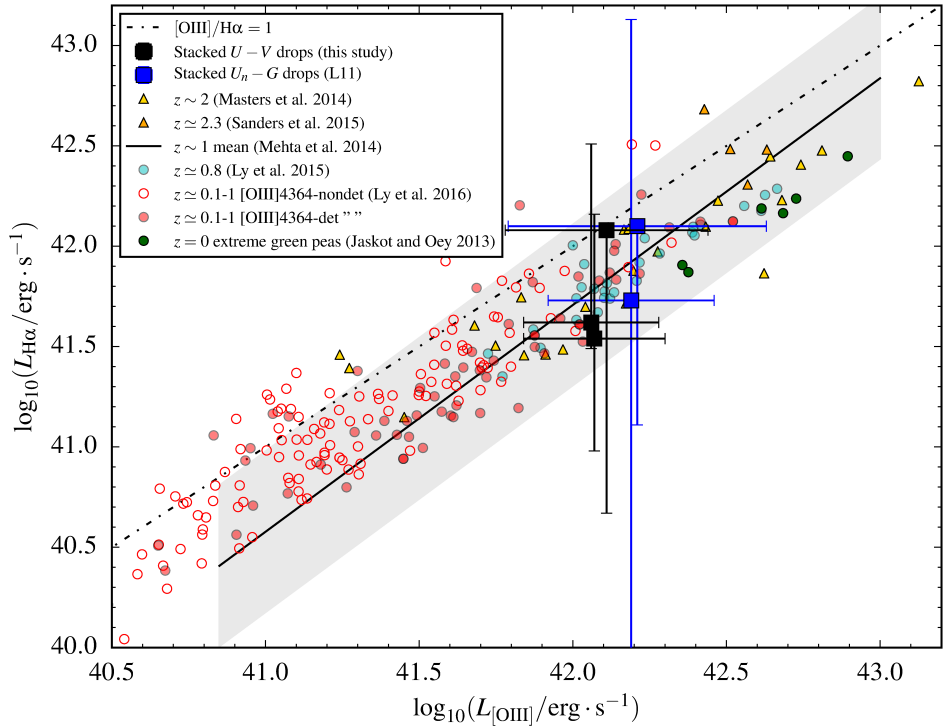}
\end{center}
\caption{ Reddened $L_{\mathrm{H}\alpha}$ from the LBG SFRs versus $L_{\mathrm{[OIII]\lambda\lambda4959,5007}}$ derived from their implied equivalent widths, compared to other samples of emission-line galaxies (see Section \ref{subsubsec:OIIIBalmer} for description of samples). 
\label{fig:LOIII} }
\end{figure*}

  The physical conditions of star-forming regions at high redshift differ significantly from local systems. An observable consequence of these differences is higher [OIII]/H$\beta$ and [OIII]/H$\alpha$ ratios measured in distant versus local galaxies \citep[e.g.,][]{teplitz2004,erb2006,shapley2005b,ly2007b,brinchmann2008,liu2008,schenker2013,steidel2014,sanders2015,holden2016}. To place the LBGs studied here in this context, we estimate [OIII]/H$\alpha$ ratio using the implied 
  $\rm{EW([OIII]\lambda\lambda4959,5007)}$ and best-fit SFRs of the stacked LBGs. The [OIII] luminosity is calculated from 
  \begin{equation}
  L_{\mathrm{[OIII]}} = 4\pi D_{\mathrm{L}}^2 f_{2.2\mu\mathrm{m,cont}} \times \mathrm{EW([OIII])},
  \end{equation}
   where $D_{\mathrm{L}}(z=3.1) = 26$ Gpc is the luminosity distance at $z=3.1$, $f_{2.2\mu\mathrm{m,cont}}$ is the continuum flux density at 2.2 $\mu$m in ergs s$^{-1}$ cm$^{-2}$ {\AA}$^{-1}$ (measured from the best-fit stellar continuum models as described above), and the equivalent widths are in the observed frame.

   We estimate the $L_{\mathrm{H}\alpha}$ of the LBGs from their model SEDs with the same approach used in \citet{ly2012}. We use the \citet{kennicutt1998} relation to calculate intrinsic H$\alpha$ luminosity from the best-fit SFRs: $L_{\mathrm{H}\alpha,\mathrm{int}} = 1.26 \times 10^{41} \cdot\mathrm{SFR}$ [M$_\odot$ yr$^{-1}]$. Then, in order to compare our LBGs with other galaxy samples, we predict their {\it reddened} H$\alpha$ luminosities by estimating the extinction for the H$\alpha$ line, $A(\mathrm{H}\alpha)$, from their best-fit visual extinctions $A_V$. The H$\alpha$ extinction can be written in terms of an extinction curve evaluated at $\lambda = 6563$ {\AA}, $k(6563\mathrm{\AA})$, 
   and the color excess for the gas $E(B-V)_{\mathrm{gas}}$ :
     \begin{equation}
  A(\mathrm{H}\alpha) = k(6563\mathrm{\AA}) \times E(B-V)_{\mathrm{gas}}.
  \end{equation} 
  For simplicity, we adopt $k(6563\mathrm{\AA})=2.00$ from the SMC extinction curve \citep{gordon2003}. We relate the color excess
  for the gas to that for the stellar continuum by assuming the \citet{calzetti2000} relation: $E(B-V)_{\mathrm{gas}} = 2.27 E(B-V)_{\mathrm{stars}}$. We note that previous studies generally support the assumption that the extinction for the gas is roughly twice that of the stars \citep[e.g.,][]{ly2012,reddy2015}. Thus:
    \begin{eqnarray}
  A(\mathrm{H}\alpha) &&= k(6563\mathrm{\AA}) \times 2.27 \times E(B-V)_{\mathrm{stars}}\\
    A(\mathrm{H}\alpha) &&= 4.54 \times E(B-V)_{\mathrm{stars}}. \nonumber
  \end{eqnarray} 
   The color excess for the stars is calculated assuming the Calzetti attenuation law: $E(B-V)_{\mathrm{stars}} = A_V / 4.05$. We note that regardless of our assumptions for the reddening of H$\alpha$, the results discussed below are not significantly impacted ($\lesssim 0.1$ dex difference). 
%  
%   
%   Then, in order to compare our LBGs with other galaxy samples, we predicted their {\it reddened} H$\alpha$ luminosities using the values of $A_V$ from the SED fits and the Calzetti attenuation law for the stars: $E(B-V)_{\mathrm{cont}} = A_V / 4.05$. Although uncertain for high redshift systems, we assume the \citet{calzetti2000} relation between extinction for the gas and for the stars: $E(B-V)_{\mathrm{gas}} = 2.27 E(B-V)_{\mathrm{cont}}$. We note that previous studies generally support the assumed relation that the extinction for the gas is roughly twice that of the stars \citep[e.g.,][]{ly2012,reddy2015}. Thus the attenuation at $\lambda=6563$ {\AA} is:
%    \begin{eqnarray}
%  A(\mathrm{H}\alpha) = k(6563\mathrm{\AA}) \times E(B-V)_{\mathrm{gas}} \\
%  = k(6563\mathrm{\AA}) \times 2.27 \times E(B-V)_{\mathrm{cont}}. \nonumber
%  \end{eqnarray} 
%  Finally, we assume $k(6563\mathrm{\AA}) = 2.00$ from the SMC extinction curve \citep{gordon2003}:
%  \begin{equation}
%  A(\mathrm{H}\alpha) = 4.54 \times E(B-V)_{\mathrm{cont}}.
%  \end{equation}
%    We note that regardless of our assumptions for the reddening of H$\alpha$, the results discussed below are not significantly altered ($\lesssim 0.1$ dex difference). 

  We plot estimated $L_{\mathrm{H}\alpha}$ versus $L_{\mathrm{[OIII]}}$ for the average LBG SEDs in Figure \ref{fig:LOIII}. The black dashed-dotted line indicates a slope of unity ([OIII]/H$\alpha$ = 1). The average $\langle i' \rangle = 25$, 26, and 27 stacked LBGs have [OIII]/H$\alpha \simeq 1.6$, 3.1, and 4.1, respectively. The solid black line and gray band represent the mean relation and scatter for WISP galaxies in the redshift range $z=0.8-1.2$ reported in \citet{mehta2015}.  Our two faintest stacked SEDs lie toward the bottom edge of the scatter, comparable to the more extreme WISP galaxies observed. Likewise, [OIII]/H$\alpha$ in the faint LBGs is consistent with the ratios observed for a subset of WISP galaxies with Magellan/FIRE spectra in \citet{masters2014}, which reach [OIII]/H$\alpha \simeq 3$ (shown on the plot as yellow triangles). 
 \citet{dominguez2013} reports line ratios for stacked WISP spectra reaching an average $\mathrm{[OIII]/H}\alpha \sim 2$  for their faintest bin of $L_{\mathrm{H}\alpha}$. At $z\sim 2.3$, galaxies from the MOSDEF survey have stacked spectra (binned by mass) indicating a maximum average [O{\sc iii}]/H$\alpha$ of $\simeq 1.8$ \citep{sanders2015}. However, these stacks (shown as orange triangles in the plot) were normalized by $L_{\mathrm{H}\alpha}$, which prevents low-metallicity objects with high [O{\sc iii}]/H$\beta$ from dominating the stacked line ratios. The most extreme galaxies at $z\sim 1.95-2.65$ in the KBSS survey \citep{steidel2014} have individual spectra that reach up to [OIII]/H$\alpha \sim 4$. The $z \lesssim 1$ ELGs from \citet{ly2015} and \citet{ly2016}, also included in Figure \ref{fig:LOIII}, form a locus with a shallower slope than the faint stacked LBGs. Perhaps most comparable to our sample is a subset of six ``extreme" green peas \citep{jaskot2013} that have $\mathrm{[OIII]/H}\alpha \simeq 3.0$ on average (shown as the green circles in the plot). 
  
 The average faint galaxy in our sample is clearly efficient at converting ionizing photons from starlight into [OIII] emission. By integrating the stellar continuum models fit to the faint average LBGs, we estimate the ratio of flux in [OIII] to observed (reddened) stellar luminosity. The ratio is $\simeq 1$\% for $\langle i' \rangle = 26$ and $\simeq 3$\% for $\langle i' \rangle = 27$ stacked LBGs.
Spectroscopic surveys of high-redshift galaxies such as those described above support this trend of high-level [OIII] emission in the distant universe. Theoretical explanations include the possibility that the interstellar medium (ISM) has a larger ionization parameter (ionizing photon density to gas density), higher electron density,  harder ionizing radiation spectrum, and/or very low metallicity \citep[e.g.,][]{kewley2013a,kewley2013b,steidel2014,nakajima2014,sanders2016}. 
%For example, the lower metallicity at a given stellar mass in high-redshift versus local galaxies, in conjunction with the anti-correlation between ionization parameter and metallicity, might be the primary factor behind a larger ionization parameter found for star-forming galaxies at high-redshift \citep{sanders2016}. 
 The very strong [OIII] emission implied for the {\it average} faint LBG in this study suggests that extreme ionization conditions (relative to those found in  most local galaxies) are ubiquitous in low-mass galaxies at $z\sim3$. Recent studies have shown that green peas (local strong [OIII]-emitters), which have ISM properties similar to LBGs, are good candidates for leaking a significant amount of ionizing radiation \citep[e.g.][]{nakajima2014,izotov2016a,izotov2016b}.
  Thus, galaxies at $z>6$ that are directly analogous to the faint LBGs in this study might have been responsible for the bulk of the cosmic reionization of the intergalactic medium. Confirmation will require deep spectroscopy at wavelengths of 3.5 $\mu$m or longer.  
This will be possible with the {\it James Webb Space Telescope}, in targeted surveys, or serendipitously with slitless spectroscopy.

%---------------------------------------------------------------------
%---------------------------------------------------------------------

\section{LBG CLUSTERING IN THE SDF} \label{sec:clustering}

\begin{figure}
\begin{center}
\includegraphics[width=0.48\textwidth]{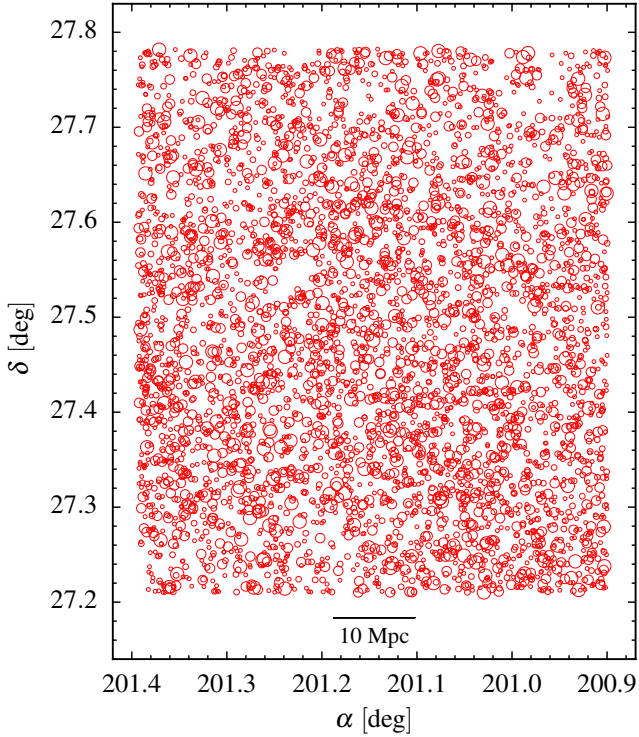}
\end{center}
\caption{%
Sky distribution of the LBGs in our sample.
%The region outlined by the thick line indicates the area observed
%in the $J$ and $K$ bands,
%while the rest is the area observed only in optical bands,
%after removal of low-quality regions.
The red circles ordered from largest to smallest represent LBGs with $23.0<R_c \le 24.5$,
 $24.5< R_c \le 25.5$, and $25.5 < R_c \le 26.5$, and $26.5 < R_c \le 27.5$, respectively.
%LBGs having $K$ magnitudes are further distinguished
%by $z'$ and $K$ magnitudes with different colors,
%where red, green, cyan, and blue represent LBGs with
%$23.0<z'\le 24.5$ and $K\le 23.46$, $24.5<z'\le 25.5$ and $K\le 23.46$,
%$23.0<z'\le 24.5$ and $K>23.46$, $24.5<z'\le 25.5$ and $K>23.46$,
%respectively.
The projected comoving scale of 10 Mpc at $z=3.3$
is shown in the bottom center.
North is up and east is to the left in this image.
\label{fig:skymap}}
\end{figure}

Clustering analysis requires highly uniform sensitivity
over a large contiguous area, because fluctuations of sensitivity can produce
spurious clustering signals
and bias the measurements of clustering strength.
We examined the sensitivity variation over the images
by dividing them into small grids and estimating the sky noise
in each of the meshes.
Based on these sky-noise maps,
we carefully defined a high-quality region
in which the sensitivity is good and uniform,
trimming the edges of the images where sky noise was systematically
larger due to dithering. The effective area with complete coverage in all of the six
optical bands and the $K$-band amounts to 876 arcmin$^2$,
after low-quality regions are discarded, including the edges.
\reffig{skymap} shows the sky distribution of the LBG sample.
Larger red circles denote objects that have brighter $R_c$-band magnitudes. 

\subsection{Angular Correlation Function} \label{subsec:acf}

We measure the clustering of both our LBG sample as a whole (all 5161 objects), and the brightest half of the sample (split by $R$-magnitude), in order to investigate mass-dependent clustering. We largely follow the methodology of \citet{yoshida2008} to calculate the angular correlation function (ACF), $\omega(\theta)$,
using the \citet{landy1993} estimator:
\begin{eqnarray} \label{eq:omega}
\omega(\theta) &=&
 \frac{dd(\theta)-2dr(\theta)+rr(\theta)}{rr(\theta)},
\end{eqnarray}
where $dd(\theta)$, $dr(\theta)$, and $rr(\theta)$
are the numbers of galaxy-galaxy, galaxy-random,
and random-random pairs, respectively,
with angular separations $\theta$.
In creating the random catalog, we avoided regions where galaxies
are not detected (for example, near bright stars). 
We generated a factor of 100 more random coordinates than the number of
galaxies in the sample, and normalized $dd$, $dr$, and $rr$ to the total number of pairs. 
Because the random pairs are subject to the same limitations as the real data,
the deficiency of faint galaxies we found within several arcseconds of other
galaxies due to confusion should not impact our estimated clustering. 

We evaluate errors and covariance matrices of both the full sample and bright-half sample by the delete-one jackknife resampling method. 
The entire survey field is divided into $36$ subfields and ACFs are calculated by omitting one subfield, repeating this procedure $36$ times. 
The covariance matrix, $C_{ij}$, is derived as
\begin{equation}
C_{ij} = \frac{N-1}{N} \sum_{k=1}^{N} \left(\omega_{k}\left(\theta_{i} \right) - {\bar \omega}\left(\theta_{i} \right) \right) \times \left(\omega_{k}\left(\theta_{j} \right) - {\bar \omega}\left(\theta_{j} \right) \right), 
\label{eq:cov}
\end{equation}
where ${\bar \omega}\left(\theta_{i} \right)$ is the averaged ACF of $i$th angular bin. 

The observed ACFs and associated error bars for the bright and full samples of LBGs are shown in Figure \ref{fig:HOD}.

\subsection{Halo Occupation Distribution}\label{subsec:hod}
\begin{figure}[tbp]
\begin{center}
\includegraphics[width=0.48\textwidth]{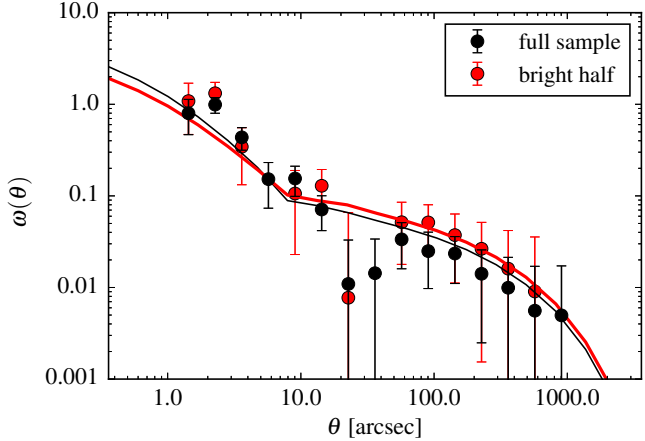}
\caption{Observed ACFs for the total (black) and bright (red) $U$-dropout galaxies. The solid lines represent the best-fit ACFs calculated by the HOD model. \label{fig:HOD} }
\end{center}
\end{figure}

At separations smaller than $\simeq$10{\arcsec}, corresponding to about 80 kpc at $z=3$, the observed ACFs in Figure \ref{fig:HOD} show a significant steepening. Such a steepening has also been found in previous LBG clustering studies, and is attributed to the additional clustering of $>1$ galaxies residing in a single dark matter halo \citep{hildebrandt2007,yoshida2008,bielby2013,bian2013}. Thus, rather than fit a single power-law to the ACF, as is commonly done, we carried out a halo occupation distribution \citep[HOD; e.g.,][]{berlind2002,zheng2005,vdbosch2007} analysis to interpret the relationship between the $U$-dropout galaxies and their host dark halos.
We employ the standard halo occupation model proposed by \citet{zheng2005}. 
The occupation of central galaxies, $\Nc$, is formulated as
\begin{equation}
\Nc = \frac{1}{2} \left[1 + {\rm erf}\left( \frac{\log{(\Mh)}-\log{(\Mmin)}}{\sigmalogM} \right) \right], 
\label{eq:Nc}
\end{equation}
whereas the occupation of satellite galaxies, $\Ns$, can be described by:
\begin{equation}
\Ns = \left( \frac{\Mh - \Mzero}{\Mone} \right)^{\alpha}. 
\label{eq:Ns}
\end{equation}
The number of total galaxies within dark halos with mass $\Mh$ is:
\begin{equation}
N(\Mh) = \Nc \left[1 + \Ns \right]. 
\label{eq:Ntotal}
\end{equation}
We vary all of the HOD free parameters, $\Mmin$, $\Mone$, $\Mzero$, $\sigmalogM$, and $\alpha$, and find the best-fit parameters to the observed ACFs. 

The HOD analysis requires assuming some analytical models to characterize dark halos. 
We use the halo mass function of \citet{sheth1999}, the radial density profile of \citet{navarro1997},  the halo bias model of \citet{tinker2010}, and the halo mass--concentration parameter relation of \citet{takada2003}. 
The redshift completeness functions $p(m,z)$ (Section \ref{subsec:lbg-comp}) are employed as the redshift distributions of each subsample. 

%We evaluate errors and covariance matrices of each magnitude subsample by the delete-one jackknife resampling method. 
%The entire survey field is divided into $36$ subfields and ACFs are calculated by omitting one subfield, repeating this procedure $36$ times. 
%The covariance matrix, $C_{ij}$, is derived as
%\begin{equation}
%C_{ij} = \frac{N-1}{N} \sum_{k=1}^{N} \left(\omega_{k}\left(\theta_{i} \right) - {\bar \omega}\left(\theta_{i} \right) \right) \times \left(\omega_{k}\left(\theta_{j} \right) - {\bar \omega}\left(\theta_{j} \right) \right), 
%\label{eq:cov}
%\end{equation}
%where ${\bar \omega}\left(\theta_{i} \right)$ is the averaged ACF of $i$-th angular bin. 
The HOD parameters are constrained through minimizing the $\chi^{2}$ as follows: 
\begin{equation}
\begin{aligned}
\chi^{2} =& \sum_{i, j} \Bigl( \omega_{{\rm obs}} (\theta_{i}) - \omega_{{\rm HOD}} (\theta_{i}) \Bigr) (C^{-1})_{ij} \Bigl( \omega_{{\rm obs}} (\theta_{j}) - \omega_{{\rm HOD}} (\theta_{j}) \Bigr) \\
&+ \frac{[n^{{\rm obs}}_{g} - n^{{\rm HOD}}_{g}]^{2}}{\sigma^{2}_{n_{g}}},
\label{eq:chi}
\end{aligned}
\end{equation}
where $C^{-1}$ is the inverse covariance matrix (Equation \ref{eq:cov}), $n^{{\rm obs}}_{g}$ and $n^{{\rm HOD}}_{g}$ are the galaxy number densities from observation and the HOD model, and $\sigma_{n_{g}}$ is the statistical $1\sigma$ error of $n^{{\rm obs}}_{g}$, respectively. 
The best-fit parameter sets and those $1\sigma$ confidence intervals are estimated by Markov Chain Monte Carlo simulation. 
We note that effects of contaminations on observed ACFs are corrected by multiplying the factor of $1/(1-f_{{\rm c}})^{2}$, where $f_{{\rm c}}$ represents the fraction of contamination. 

The best-fit ACFs for the full and bright LBG samples, computed by the HOD modeling, are shown with the observed ACFs in Figure~\ref{fig:HOD}. 
The best-fit HOD parameters and deduced parameters, i.e., mean halo masses and satellite fractions, are listed in Table~\ref{tab:HODparams}. 
Mean halo masses, $\Mhalo$, and satellite fractions, $\fsat$, are defined as
\begin{equation}
\Mhalo = \frac{\int d\Mh \Mh n(\Mh) N(\Mh)}{\int d\Mh n(\Mh) N(\Mh)}, 
\label{eq:Mhalo}
\end{equation}
and 
\begin{equation}
\fsat = 1 - \frac{\int d\Mh n(\Mh) \Nc}{\int d\Mh n(\Mh) N(\Mh)}, 
\label{eq:fsat}
\end{equation}
where $n(\Mh)$ is the halo mass function. 

The mean halo masses implied for the total LBG sample and for the bright-half subsample are $\Mhalo = (1.9^{+0.6}_{-0.5}) \times 10^{11}$ $h^{-1} M_{{\odot}}$ and $\Mhalo = (3.1^{+0.8}_{-0.6}) \times 10^{11}$ $h^{-1} M_{{\odot}}$, respectively. \citet{bian2013} have also carried out HOD analysis for their $z\sim 3$ LBG sample in two magnitude bins and found $\Mhalo = (2.5 \pm 0.3) \times 10^{12}$ $h^{-1} M_{{\odot}}$ for LBGs with $24.0 < R < 24.5$, and $\Mhalo = (3.3^{+0.6}_{-0.4}) \times 10^{12}$ $h^{-1} M_{{\odot}}$ for $23.5 < R < 24.0$. The mean halo masses derived for our sample of LBGs are significantly smaller, consistent with the fact that we include much fainter galaxies in the sample (down to $R \simeq 28$ for the full sample). Overall, the HOD analysis of our LBG sample and comparison to that of brighter samples support the notion that more massive dark matter halos host more luminous LBGs. 

%\newpage
\begin{table*}[htb]
\begin{center}
\caption{Best-fit HOD parameters and deduced parameters with $1\sigma$ confidence intervals \label{tab:HODparams}}
\begin{tabular}{lcccccccc} \hline \hline
 $N$ & $\log_{10}(\Mmin/h^{-1} \Msun)$ & $\log_{10}(\Mone/h^{-1} \Msun)$ & $\log_{10}(\Mzero/h^{-1} \Msun)$ & $\sigmalogM$ & $\alpha$ & $\log_{10}(\Mhalo/h^{-1} \Msun)$ & $\fsat$ & $\chi^{2}/{\rm dof}$ \\ \hline
 $5161$ & $10.94^{+0.13}_{-0.28}$ & $12.75^{+0.48}_{-0.34}$ & $8.18^{+2.25}_{-2.16}$ & $0.51^{+0.18}_{-0.32}$ & $0.81^{+0.38}_{-0.20}$ & $11.29 \pm 0.12$ & $0.05 \pm 0.03$ & $0.61$ \\ 
$2581$ & $11.21^{+0.11}_{-0.12}$ & $13.20^{+0.50}_{-0.42}$ & $8.18^{+2.22}_{-2.18}$ & $0.49^{+0.21}_{-0.29}$ & $0.83^{+0.38}_{-0.19}$ & $11.49 \pm 0.10$ & $0.03 \pm 0.02$ & $1.40$ \\ \hline
\end{tabular}
\end{center}
\end{table*}

%---------------------------------------------------------------------
%---------------------------------------------------------------------
\section{SUMMARY} \label{sec:sum}

Using deep multi-waveband imaging data
from optical to infrared wavelengths
in the Subaru Deep Field, we investigate the 
LF, physical properties, and clustering of 
a large sample of Lyman-break galaxies
(LBGs) at $z\sim 3$. The LBGs are selected by $U-V$ and $V-R_c$ colors in 
one contiguous area
of 876 arcmin$^2$ down to $R_c=27.8$, yielding a sample of 5161
LBGs in total. A subset of 140 of these LBGs are detected 
in near-infrared wavelengths (in both $H$ and $K$ band). 
%The total number of LBG candidates detected is 5161,
%about half of which 
%are found in the region observed deeply in the $H$ and $K$ bands.
We use Monte Carlo simulations to estimate the redshift distribution
function and the fraction of contamination by interlopers of the
LBG samples.  As expected, our LBG search is fairly uniformly sensitive
to redshifts between $z=2.9$ and 3.5, with a declining tail to higher redshifts.

Using our completeness simulations, we calculate the LBG LF and find
that our results are broadly consistent with previous LF determinations at
$z\sim 3$. We fit our LBG LF with a standard Schechter function, deriving a 
steep faint end slope of $\alpha=-1.78 \pm 0.05$ and a characteristic magnitude of $M_{\mathrm{UV}}^* = -20.86\pm 0.11$.
We also measure clustering for the LBG sample and model the angular correlation function using the halo occupation distribution framework. We find that, on average, the bright half of the LBG sample resides in more massive dark matter halos than the sample as a whole. This suggests that more-luminous LBGs (which host a larger star formation rate) reside in more-massive dark matter halos.

To infer physical properties of the LBGs, we construct their average rest-frame UV-to-NIR (observed optical to mid-IR) SEDs. 
The SEDs are generated by binning our sample according to $i'$ magnitude, and obtaining stacked LBG detections in the infrared by 
median-averaging at the optical positions of the LBGs.  
Stacking is performed in WFCAM $J$, $H$, and $K$ bands through IRAC [3.6], [4.5], [5.8], and [8.0] $\mu$m filters. 
In the stacks of faint LBGs, we find a background depression in the immediate vicinity of the stacked object. 
We confirm that this background deficit is due to the source detection algorithm, SExtractor, preferentially selecting only 
faint galaxies that happen to fall in regions of low background counts (they are relatively isolated in the nearly confusion-limited data). 
We suggest that most (if not all) catalogs generated from similar detection routines suffer from this bias.  
Thus, stacks of objects in such catalogs should be appropriately corrected for the locally faint background. 

We applied these corrections to all faint stacked images 
prior to measuring fluxes. 
The average LBG SEDs are then formed by combining the median $UBVR_ci'z'$ photometry with photometry from the corrected stacked images. 
We fit the average SEDs with stellar population synthesis templates and infer stellar mass and SFRs, among other properties. 
The average LBGs range in stellar mass from $\simeq 10^{10}$ $\mathrm{M}_\odot$ down to $\simeq 10^8$ $\mathrm{M}_\odot$, 
and are forming stars at rates of $50$ $\mathrm{M}_\odot$ yr$^{-1}$ to 3 $\mathrm{M}_\odot$ yr$^{-1}$, from bins of $\langle i' \rangle = 24$ to $\langle i' \rangle = 27$, respectively. 
The properties are generally consistent with other samples of LBGs at this redshift and lie close to the ``star-forming main sequence" at $z \sim 3$.  

The average SEDs for the faint, low-mass bins have additional 
features that are indicative of strong nebular emission. 
There is a large excess in the $K$-band flux that we attribute to the contribution of [OIII]$\lambda\lambda$4959,5007+H$\beta$ emission. 
From the excess, we estimate rest-frame equivalent widths reaching $\rm{EW}_0(\rm{[OIII]}) \gtrsim 1000$ {\AA} 
for the faintest magnitude bin ($\langle i' \rangle = 27$). 
This result suggests that the average low-mass galaxy that is forming stars at $z\sim 3$ radiates a large fraction 
of its power,  $\sim 1$\% or more of the total stellar luminosity, in this [OIII] doublet. Furthermore, this efficiency in [OIII] emission 
increases as galaxy stellar mass decreases and sSFR increases. Strongly star-forming dwarf galaxies
can thus be detected by the excess brightness produced by  [OIII]$\lambda\lambda$4959,5007 in their broad-band SEDs, even if they are too faint
for spectroscopy.
 
The faint average LBGs are comparable to the most extreme emission-line galaxies at lower redshift. 
The result appears plausible in the context of emerging evidence for ubiquitous strong [OIII] emission in the high-$z$ universe. 
Recently, there have been discoveries of significant leakage of ionizing radiation from green pea galaxies, which are local analogs 
of [OIII]-emitting LBGs. We suggest that low-mass systems with strong [OIII] emission, which are seemingly pervasive in the high-redshift universe, 
are strong candidates to produce the bulk of cosmic reionization at $z>6$. 
 
%Finally, we measure the clustering strength of the LBGs using the 2-point correlation function. 
%The amplitude of the 2-point correlation function for LBGs steepens for separations
%smaller than $10\arcsec$, corresponding to 77 kpc (projected) at $z=3$.  
%This is likely the signature of non-linear effects, such as dark haloes 
%being populated by more than one galaxy.
%The correlation length is found to be $r_0 = 4.88$ $h^{-1}$Mpc, corresponding to a bias factor with respect to dark matter of $b=2.8$, and implying that the LBG populations reside in dark matter halos of total mass $M_{DM} = 1.6 \times 10^{12}$. 
 
%The correlation length is found to be $r_0 = 4.88h^{-1}$  kpc
%for the total sample and $r_0 = 8.02h^{-1}$ kpc for our subsample of massive (NIR-detected)
%LBGs above $L_*$. These correspond to bias factors with respect to dark matter of $b=2.8$
%and $b=4.8$, respectively, implying that the LBG populations reside in dark matter halos of
%total mass $M_{DM} = 1.6 \times 10^{12}$ $M_{\odot}$ and $M_{DM} = 7.1 \times 10^{12}$ $M_{\odot}$,
%respectively. 

This study of 5161 $U-V$ dropouts LBGs yields some results 
consistent with other studies of star-forming galaxies at $z\sim 3$.
However, we were surprised to find such strong and widespread [OIII] emission in the average (faint, i.e., typical) LBGs.
Telescopes like the {\it Wide-Field Infrared Survey Telescope} and the {\it James Webb Space Telescope}, will 
undoubtedly utilize strong [OIII] emission lines to characterize the typical galaxies at $z > 3$ \citep{colbert2013}.  
This doublet could prove at least as important as Ly$\alpha$,  or even more so,  in studying those galaxies
that likely re-ionized the universe. 
One positive conclusion is that, as the infrared spectroscopic searches push deeper, they will benefit
more and more from the relative ease of detecting [OIII] lines in the fainter galaxies.
One concern is that surveys of large-scale structure based on [OIII] line emission will be strongly biased
toward the least massive galaxies, which may have relatively weaker clustering.

%---------------------------------------------------------------------
%---------------------------------------------------------------------
\acknowledgments
We would like to thank the referee for valuable comments and thoughtful
suggestions. 
We are deeply grateful to the Subaru and Mayall Telescope staffs for their
invaluable support in the observations at Mauna Kea and Kitt Peak
National Observatory.
We also thank the SDF collaboration for their contribution
to this research.
M.M. acknowledges support from the Japan Society for the Promotion of
Science (JSPS), and the Visitor grant program of the National Astronomical
Observatory of Japan.
%through JSPS Research Fellowship for Young Scientists.

This work is based in part on observations made with the {\it Spitzer
Space Telescope}, obtained from the NASA/ IPAC Infrared Science
Archive, both of which are operated by the Jet Propulsion Laboratory, 
California Institute of Technology under a contract with the National Aeronautics and Space Administration.

Chun Ly was supported by an appointment to the NASA Postdoctoral Program at the Goddard Space Flight Center, administered by Oak Ridge Associated Universities and Universities Space Research Association through contracts with NASA, and by NASA Astrophysics Data Analysis Program grant NNH14ZDA001N.  N. Kashikawa acknowledges support from the JSPS, KAKNHI grant number 15H03645.

% --------- facilities 
{\it Facilities: } \facility{Subaru (Suprime-Cam)}, \facility{Mayall (MOSAIC-1)},
 \facility{UKIRT (WFCAM)}, \facility{Mayall (NEWFIRM)}, \facility{{\it Spitzer} (IRAC)}

%---------------------------------------------------------------------
% Bibliography
\bibliographystyle{apj}
\tracingmacros=2

\bibliography{sdfudropsbib.bib}

%--------- Appendix-------------------
% USED TO BE CLUSTERING SECTION, BUT CLUSTERING NOW MOVED TO MAIN BODY AS 
% LAST SECTION
%\newpage
%\begin{landscape}
%\begin{table*}[htb]
%\begin{center}
%\caption{Best-fit HOD parameters and deduced parameters with $1\sigma$ confidence intervals \label{tab:HODparams}}
%\begin{tabular}{ccccccccc} \hline \hline
% $N$ & $\log_{10}(\Mmin/h^{-1} \Msun)$ & $\log_{10}(\Mone/h^{-1} \Msun)$ & $\log_{10}(\Mzero/h^{-1} \Msun)$ & $\sigmalogM$ & $\alpha$ & $\log_{10}(\Mhalo/h^{-1} \Msun)$ & $\fsat$ & $\chi^{2}/{\rm dof}$ \\ \hline
% $5,161$ & $10.94^{+0.13}_{-0.28}$ & $12.75^{+0.48}_{-0.34}$ & $8.18^{+2.25}_{-2.16}$ & $0.51^{+0.18}_{-0.32}$ & $0.81^{+0.38}_{-0.20}$ & $11.29 \pm 0.12$ & $0.05 \pm 0.03$ & $0.61$ \\ 
%$2,581$ & $11.21^{+0.11}_{-0.12}$ & $13.20^{+0.50}_{-0.42}$ & $8.18^{+2.22}_{-2.18}$ & $0.49^{+0.21}_{-0.29}$ & $0.83^{+0.38}_{-0.19}$ & $11.49 \pm 0.10$ & $0.03 \pm 0.02$ & $1.40$ \\ \hline
%\end{tabular}
%\end{center}
%\end{table*}
%\end{landscape}
%
%

\end{document}